\documentclass[acmsmall,screen,nonacm]{acmart}

\usepackage{balance}
\usepackage{xspace}
\usepackage{subfig}
\usepackage{enumitem} 
\usepackage{ulem} 
\usepackage{array} 
\usepackage[noabbrev, capitalise]{cleveref}
\usepackage{wrapfig}
\usepackage{multirow} 

\usepackage{caption}

\newcolumntype{C}[1]{>{\centering\arraybackslash}p{#1}}

\definecolor{darkgreen}{rgb}{0.0, 0.4, 0.4}

\newcommand{\name}{PEMI\xspace}

\newcommand{\smalltitle}[1]{
  \medskip 
  \noindent\textbf{#1}
}

\AtBeginDocument{
  }

\setcopyright{none}
\copyrightyear{2018}
\acmYear{2018}
\acmDOI{XXXXXXX.XXXXXXX}

\acmConference[Conference acronym 'XX]{Make sure to enter the correct
  conference title from your rights confirmation emai}{June 03--05,
  2018}{Woodstock, NY}
\acmISBN{978-1-4503-XXXX-X/18/06}

\renewcommand\footnotetextcopyrightpermission[1]{}
\begin{document}

\title{\name: Transparent Performance Enhancements for QUIC}

\author{Jie Zhang}
\affiliation{
  \institution{Tsinghua University}
  \city{Beijing}
  \country{China}
  }
\email{zhangjie19thu@gmail.com}

\author{Lei Zhang}
\affiliation{
  \institution{Tsinghua University}
  \city{Beijing}
  \country{China}
  }
\email{zhanglei@zgclab.edu.cn}

\author{Ziyi Wang}
\affiliation{
  \institution{Beijing University of Posts and Telecommunications}
  \city{Beijing}
  \country{China}
  }
\email{wangziyi0821@gmail.com}

\author{Chenxiang Sun}
\affiliation{
  \institution{Tsinghua University}
  \city{Beijing}
  \country{China}
  }
\email{scx24@mails.tsinghua.edu.cn}

\author{Yuming Hu}
\affiliation{
  \institution{University of Minnesota -- Twin Cities}
  \city{Minneapolis}
  \state{Minnesota}
  \country{USA}
  }
\email{yuming.hum@gmail.com}

\author{Xiaohui Xie}
\affiliation{
  \institution{Tsinghua University}
  \city{Beijing}
  \country{China}
  }
\email{xiexiaohui@tsinghua.edu.cn}

\author{Zeqi Lai}
\affiliation{
  \institution{Tsinghua University}
  \city{Beijing}
  \country{China}
  }
\email{zeqilai@tsinghua.edu.cn}

\author{Yong Cui}
\affiliation{
  \institution{Tsinghua University}
  \city{Beijing}
  \country{China}
  }
\email{cuiyong@tsinghua.edu.cn}

\begin{abstract}
  QUIC, as the transport layer of the next-generation Web stack (HTTP/3), natively provides security and performance improvements over TCP-based stacks. 
  However, since QUIC provides end-to-end encryption for both data and packet headers, in-network assistance like Performance-Enhancing Proxy (PEP) is unavailable for QUIC. To achieve the similar optimization as TCP, some works seek to collaborate endpoints and middleboxes to provide in-network assistance for QUIC.
  But involving both host and in-network devices increases the difficulty of deployment in the Internet.

  In this paper, by analyzing the QUIC standard, implementations, and the locality of application traffic, we identify opportunities for transparent middleboxes to measure RTT and infer packet loss for QUIC connections, despite the absence of plaintext ACK information.
  We then propose \name as a concrete system that continuously measures RTT and infers lost packets, enabling fast retransmissions for QUIC. 
  \name enables performance enhancement for QUIC in a completely transparent manner, without requiring any explicit cooperation from the endpoints.
  To keep fairness, \name employs a delay-based congestion control and utilizes feedback-based methods to enforce CWND. 
  Extensive evaluation results, including Mininet and trace-driven dynamic experiments, show that \name can significantly improve the performance of QUIC. For example, in the Mininet experiments, \name increases the goodput of file transfers by up to 2.5$\times$, and reduces the 90th percentile jitter of RTC frames by 20-75\%.
  
\end{abstract}



\maketitle
\section{Introduction}

Traditionally, TCP~\cite{tcp1988sigcomm} and the applications above it adhere to the end-to-end principle, running on hosts and leaving processing across network devices to IP and lower-layer protocols.
However, an end-to-end connection may experience significant performance degradation if it crosses various types of network links~\cite{rfc3135pep, rfc3449pathasymmetry}, such as the combination of wired, Wi-Fi, cellular, and satellite networks.
With the increasing prevalence of different networks, in-network assistance like Performance-Enhancement Proxy(PEP)~\cite{rfc3135pep} has received extensive research and deployment~\cite{snoop_pep1995mobicom, milliproxy2017,pepsal2006, pepsatellite2005infocom, peplte2012, pep2019mmwave}.
PEPs are deployed on middleboxes to optimize the performance of TCP, by transparently intercepting TCP connections and mimicking communications with both endpoints. By creating these split connections, PEP can perform fast packet retransmissions on the split connection with lower latency and provide congestion control tailored to the sub-link.
Studies have shown that over 25\% of the paths on the Internet exhibit PEP-associated behaviors, such as TCP splitting and proxy-like functions~\cite{edeline2019, imc2011tcpextend,sidekick2024nsdi}.

PEPs depend on the plaintext information in TCP headers to split the connections and perform optimizations. 
The TCP header contains fields like SYN, sequence number, and acknowledgment (ACK).
These allow PEP to ascertain the TCP connection state and infer packet loss in ACK-based manners similar to end-to-end protocols. The plaintext header also enables PEP to fully proxy the TCP connection, thereby establishing new congestion control instances that it can directly manipulate.
However, deploying functions that rely on fully plaintext packet headers on middleboxes leads to ossification problems of the network, making it difficult to extend TCP~\cite{imc2011tcpextend}. 

New proposed security protocols like QUIC, encrypt the vast majority of header information to avoid ossification~\cite{quic_sigcomm_2017}.
Since the essential information for PEPs is invisible, it is impossible to deploy PEPs for QUIC connections.
This hinders the in-network assistance for QUIC, resulting in performance disadvantages compared to TCP in some scenarios~\cite{quic2022satcom, quicsatellite2020panrg}.

Recently, several studies have attempted to develop performance enhancements for QUIC by co-designing or collaboration of hosts and in-network devices~\cite{sidekick2024nsdi, sidecar2022hotnets,tecc2024nsdi}.
Unlike PEP, these efforts are unable to achieve transparency to the endpoints.
They rely on additional in-band~(\cite{tecc2024nsdi}) or out-of-band~(\cite{sidekick2024nsdi}) protocols for conveying information and require modifications to the QUIC implementation at the end hosts to leverage this information.
This increases the difficulty of deployment due to at least two factors. 
\textbf{First}, since modifications to the QUIC implementation are required, it necessitates changes to the QUIC standard (or at least the addition of new extensions) to make widespread QUIC support feasible. 
Moreover, the formulation of such standards needs to comprehensively consider the requirements and circumstances of QUIC and middleboxes, making the process quite challenging. 
\textbf{Second}, to benefit a connection from such schemes, it needs compatibility not only from the endpoints but also middleboxes. Isolated implementations from one participant are temporarily unusable, which could lead to a chicken-and-egg problem in deployment. That is, application providers may wish to wait for middlebox support before implementing efforts, while middlebox vendors want to wait for widespread QUIC support. This issue hinders deployment even when it involves only end-to-end protocols when both server and client require modifications~\cite{plugQUIC2019sigcomm}.

In this paper, we aim to realize the first \textit{completely transparent in-network assistance for QUIC} that: supports the standard QUIC protocol and needs no modification to the endpoint QUIC implementations; requires deployment only on middleboxes, without any out-of-band/in-band cooperation from end hosts. 
Achieving this goal is highly challenging because middleboxes cannot see information such as packet numbers and ACKs of QUIC, making it seemingly impossible to perform operations like RTT measurement and loss detection. 
Moreover, middleboxes cannot establish sub-connections that they can control directly, while it is also difficult to indirectly control the endpoint transmission due to the lack of cooperation. In summary, it is so challenging that appears to be impossible.

We propose a \textit{best-effort} solution called \textbf{\name} to accomplish such \uline{\textbf{P}}erformance \uline{\textbf{E}}nhancements which seems a "\uline{\textbf{M}}ission: \uline{\textbf{I}}mpossible".
Despite the encryption of packet numbers and ACK frames in QUIC, our analysis of the QUIC standard, implementations, and application traffic reveals structural opportunities that make transparent RTT measurement and loss detection feasible.
\name leverages the locality of packets to infer the correspondence between the sent packets and the reply packets. It divides the packets of a connection into flowlets, where a flowlet is a burst of packets separated by sufficient time gaps from other packets of the connection~\cite{flare2007sigcomm}.
\name finds possible reply packets for every flowlet, and then infers the sent-reply mapping.
\name filters inferred lost packets from unreplied packets by utilizing heuristics derived from ACK-triggering behaviors of QUIC standard and mainstream stacks.
By utilizing information from the QUIC handshake process, \name is able to measure the initial RTT to both endpoints.
It then continuously updates RTT estimates by observing sent and reply packets within flowlets.
To remain robust under highly dynamic network conditions, \name incorporates additional best-effort signals available at the middlebox, including the spin bit~\cite{rfc9000} and ICMP probing.
Using the measured RTT, \name implements a delay-based congestion control algorithm modified from Copa~\cite{copa2018nsdi}. This avoids harmful retransmissions during congestion and overly aggressive behavior on lossy links.

We implement a prototype of \name and conduct extensive emulation experiments using both Mininet~\cite{mininet} and trace-driven dynamic networks~\cite{sentosa2025cellreplay}. The results show that \name can effectively enhance the performance of QUIC on lossy links. 
For example, in Mininet tests, it achieves up to 2.5$\times$ higher goodput of file transfers compared to QUIC without \name. 
For the frame-level jitter of RTC frames, it reduces the median by 45\%-70\% and the 90th percentile by 20\%-75\% across different QUIC stacks.
Under trace-driven dynamic networks, \name also significantly reduces the tail frame-level jitter and frame delay.
\name is less aggressive than PEPs under packet loss, thereby preserving QUIC’s TCP-friendly behavior.
The design of \name introduces minimal computational overhead, accounting for less than 3\% of CPU cycles, while the remaining over 97\% of CPU cycles are dedicated to handling packet reception and forwarding, which is also required by PEPs.
\name takes the first step toward transparent performance enhancement for secure transport protocols. We release our implementation and tools at \url{https://github.com/zhjie233/pemi} to the community to facilitate further exploration~(see \cref{sec:appendix:opensource}).

\section{Background}
\subsection{Performance-Enhancement Proxy for TCP}
\begin{wrapfigure}{r}{0.6\textwidth}
    \centering
    \includegraphics[width=0.9\linewidth]{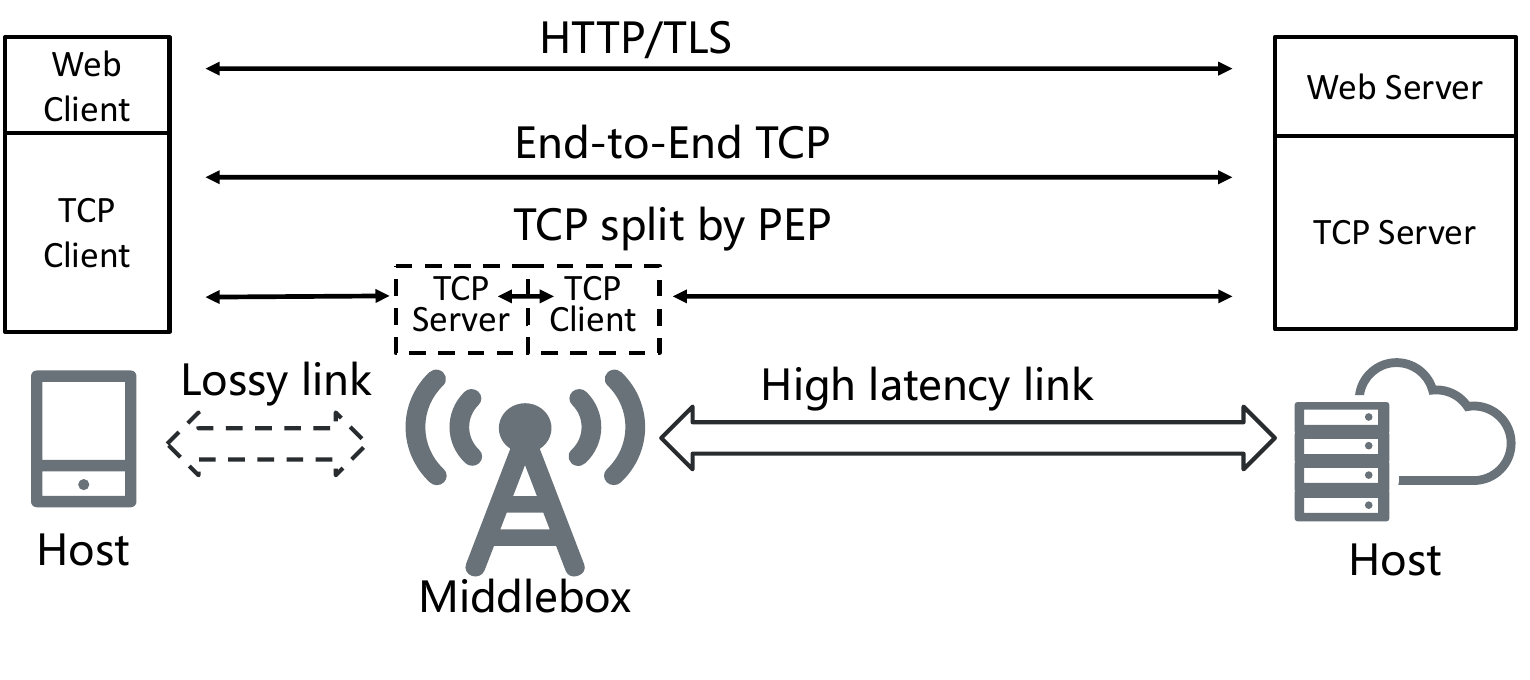}
    \caption{Sample deployment scenario of PEP.
    PEP creates a deceptive server and client, communicating with both ends.
    From the perspective of the end hosts, they are communicating with each other.
    By leveraging the advantage that sub-links have lower latency than complete end-to-end links, PEP can perform fast retransmissions when packet loss occurs in the network.}
    \Description{XXX}
    \label{fig:pep_scenario} 
\end{wrapfigure}

Performance-enhancement Proxies(PEPs)~\cite{rfc3135pep} are deployed on network middleboxes. They transparently enhance the data transmission of applications over TCP. \cref{fig:pep_scenario} shows a typical scenario where a PEP brings benefits. PEPs can be deployed on edge middleboxes, such as base stations, Wi-Fi access points, and edge servers. PEPs are able to retransmit the lost packets quickly for the lossy links~\cite{snoop_pep1995mobicom, milliproxy2017}, avoiding the end-to-end retransmission with much larger RTT. 
PEPs also benefit flows across heterogeneous links by splitting the end-to-end connection into multiple parts~\cite{pepsal2006, pepsatellite2005infocom, peplte2012, pep2019mmwave}. 
According to prior studies~\cite{edeline2019, imc2011tcpextend,sidekick2024nsdi}, 
at least 25\% of Internet paths exhibit TCP-interfering middlebox behaviors (e.g., TCP splitting and proxy-like functions), suggesting the widespread presence of in-network TCP assistance.

However, the deployment of PEPs also introduces ossification issues of TCP~\cite{De-Ossifying2017,imc2011tcpextend}.
The PEPs for TCP rely on the plaintext information in the packet headers, such as the sequence number and acknowledgment number. Some other functions on middleboxes, like NATs, firewalls, and load-balancers, also depend on the header information.
These persistent requirements in the network make it difficult to update the TCP wire format.
The ossification of TCP becomes one of the top motivations for the development of secure transport protocols with encryption of packet headers~\cite{quic_sigcomm_2017}.

\subsection{Performance-Enhancement for QUIC}
Since proposed by Google~\cite{quic_sigcomm_2017} and standardized by IETF~\cite{rfc9000}, QUIC has become the most promising protocol to replace TCP. It is the transport layer of next generation HTTP/3~\cite{rfc9114http3} and is supported by many mainstream Web browsers~\cite{chromiumquic, neqo}, servers~\cite{nginxquic}, websites~\cite{fbquic}, infrastructures~\cite{cloudflarequic}, and tools~\cite{curlhttp3}. 

QUIC encrypts most header fields, including the packet numbers and ACK frames used in loss detection~\cite{rfc9000}. It also prevents middlebox devices from splitting the connection into multiple parts. These features make it impossible to deploy TCP-like PEPs for QUIC.

Recently, a series of efforts~\cite{sidekick2024nsdi, blindbox2015sigcomm,sidecar2022hotnets,tecc2024nsdi} have been made to enhance the performance of QUIC. However, they all require the co-design or collaboration of endpoints and middleboxes, significantly increasing the deployment difficulty and complexity. This also induces the risk of protocol ossification.
In this paper, we attempt to address the following question for the first time: 

\textit{Is it possible to provide performance enhancement that is completely transparent to endpoints in the era of secure protocols?}

\section{Opportunities and Observations}
\label{sec:quicstudy}

To ensure that lost data eventually arrives at the receiver, packet loss must be recovered either by retransmission or by reconstructing it from redundancy, such as forward error correction (FEC).
However, FEC cannot be transparently deployed on middleboxes, as it requires receiver-side cooperation for decoding.
As a result, in transparent deployment scenarios, a middlebox can only rely on retransmission to recover packet losses.
The core challenge lies in identifying lost packets when both QUIC packet numbers and ACK frames are encrypted.
If such identifications were possible, a middlebox could potentially retransmit lost QUIC packets directly at the UDP layer, thereby improving end-to-end performance.

\subsection{Opportunities for Transparent Loss Recovery in QUIC}

Existing transport protocols (e.g., TCP, QUIC) and middlebox enhancements (e.g., PEPs) use ACK-based mechanisms to detect loss. 
The core of ACK-based mechanisms is that they record a unique number for every protected packet (sequence number in TCP and packet number in QUIC), and the ACK will acknowledge the numbers.
An unacknowledged packet is inferred as lost after a timeout or receiving enough reordered ACKs of later packets.
To employ this approach, the ends/middleboxes must have the ability to see the numbers of packets and the ACKs.

In a transparent deployment context, middleboxes cannot observe QUIC packet numbers or ACK frames due to encryption. 
As a result, conventional ACK-based loss detection mechanisms are no longer applicable.
This raises a fundamental question: whether packet loss can still be detected by a middlebox using only the information that remains visible in encrypted QUIC traffic.
While such inference is inherently probabilistic and limited, it reveals opportunities for middlebox-assisted loss recovery without endpoint modifications.

Specifically, even under QUIC encryption, a middlebox can still observe several types
of information: (1) information visible at the UDP layer, including packet arrival time, size, and header fields; 
(2) a limited set of fields in the QUIC header,
such as the type of long-header packets (e.g., Initial, Handshake, and Retry)~\cite{rfc9000};
(3) signals explicitly exposed to the network, such as the spin bit~\cite{rfc9000,rfc9312}; and
(4) information obtained via active probing like ICMP measurements.
Using only this visible information, a middlebox may, in some cases, infer whether a packet has been received by the peer.
For example, if a middlebox observes a QUIC Initial packet sent from host~A to host~B, followed by a Handshake packet sent from~B to~A, it can infer that the Initial packet was received by~B and that the Handshake packet constitutes a response to that Initial packet.
More generally, a middlebox may combine RTT estimates (e.g., measured via the spin bit~\cite{rfc9312}) with packet transmission timestamps to reason about packet reception. 
To build intuition, consider a simplified scenario—while real QUIC stacks exhibit more complex behavior—in which a receiver immediately responds to each received packet with a packet carrying an acknowledgment. 
In such a case, if no response is observed within an RTT-scaled time window after a packet is sent, the middlebox may infer that the packet was lost.
In practice, however, QUIC response behavior is far more complex than this simplified model. 
We therefore first examine how reply packets are triggered in QUIC, followed by an exploration of whether sent and reply packets can be continuously matched to measure RTT and infer packet loss.

\subsection{Observations from the QUIC Standard and Implementations}
\label{sec:quicobservations}

\subsubsection{ACK-Sending Features in the QUIC Standard}

The timeliness and frequency of ACK transmission represent a classic trade-off~\cite{rfc9000,rfc5681,2007reduceACK,tack2020sigcomm}.
Prompt acknowledgments enable fast reactions for ACK-driven mechanisms such as
congestion control and loss detection, but introduce transmission overhead. 
Less frequent acknowledgments can reduce overhead, but may negatively impact algorithm performance.
The QUIC standard regulates this trade-off by permitting delayed and aggregated acknowledgments while bounding their potential negative effects.

\noindent\textbf{ACK triggering frequency.}
The QUIC standard specifies that the receiver should send an ACK after receiving two or more ACK-eliciting packets\footnote{An ACK-eliciting packet is defined in~\cite{rfc9000} as a QUIC packet that contains frames other than ACK, PADDING, and CONNECTION\_CLOSE, and therefore triggers an acknowledgment at the receiver.}~\cite{rfc9000}.
This implies that a receiver is permitted to either acknowledge each received packet immediately or defer acknowledgment until another packet is received.
Furthermore, the standard allows a receiver to process multiple available packets before deciding whether to transmit an ACK, meaning that when packets are received in bursts, QUIC stacks may defer acknowledgment until several packets have been processed.

\noindent\textbf{Maximum ACK delay.}
To prevent unbounded acknowledgment deferral, QUIC specifies a maximum allowable delay for sending ACKs (denoted as \texttt{max\_ack\_delay}). This constraint ensures that even when acknowledgments are deliberately aggregated, a receiver cannot indefinitely postpone feedback to the sender.

\noindent\textbf{ACK piggybacking.}
QUIC allows ACK frames to be coalesced with application data by encapsulating ACK frames together with STREAM frames in a single QUIC packet. 
As a result, acknowledgments may be piggybacked on data packets rather than sent as standalone packets.

\subsubsection{Observations from stacks.}
\label{sec:quicackstudy}

To understand how the QUIC standard is concretely implemented in practice, we examined the ACK transmission frequency of several mainstream QUIC stacks and analyzed the source code of three of them.

\noindent\textbf{Reply-packet ratios.}
Using the QUIC Interop Runner (QIR)~\cite{seemann2020automating,quicinterop}, we analyze the five most popular\footnote{Ranked by GitHub stars as of October~2025. \texttt{haproxy} is excluded since it can operate only as a server, and \texttt{neqo} is excluded because it failed the QIR \texttt{transfer} test in our environment.} QIR-supported open-source QUIC stacks.
We run QIR’s \texttt{transfer} test and then compute the ratio between reply-direction packets and sent-direction packets. 
Each experiment is repeated twice and the average is shown.
The results are summarized in \cref{tab:quic-reply-ratios}.
It shows that the reply-packet ratio is primarily determined by the client-side QUIC stack. Moreover, the stacks clearly fall into two distinct groups: when quiche is used as the client, it generates reply packets at nearly 100\% of the number of sent packets, whereas the other four stacks as clients produce reply packets at approximately 50\%.

\begin{table}[h]
    \centering
    \caption{Reply/sent packet ratio between different QUIC stacks.}
    \begin{tabular}{c|ccccc}
        \toprule
        Server $\backslash$ Client & aioquic & msquic & quic-go & quiche & quinn \\
        \midrule
        aioquic & 49.1 \% & 42.1 \% & 50.3 \% & 101.4 \% & 50.8 \% \\
        msquic  & 50.1 \% & 50.2 \% & 53.0 \% & 101.1 \% & 51.0 \% \\
        quic-go & 49.8 \% & 50.8 \% & 51.4 \% & 101.6 \% & 51.0 \% \\
        quiche  & 45.1 \% & 51.0 \% & 51.2 \% &  99.7 \% & 50.9 \% \\
        quinn   & 50.0 \% & 50.5 \% & 51.4 \% & 101.9 \% & 51.0 \% \\
        \bottomrule
    \end{tabular}
    \label{tab:quic-reply-ratios}
\end{table}

\noindent\textbf{Source-code analysis of ACK-triggering mechanisms.}
To investigate the underlying reasons for these behaviors, we examined the source code of three QUIC stacks—quiche, quinn, and quic-go. We identified their primary ACK-triggering mechanisms through manual source-code inspection and LLM-assisted analysis.
All LLM identified mechanisms were validated via manual checking to ensure correctness.
\cref{tab:source-analyze-ack} summarizes the mechanisms we found.\footnote{Although we attempted to be comprehensive, some implemented mechanisms may still have been overlooked. }
The source-code inspection shows that quiche does not have more ACK-triggering mechanisms than the other stacks—in fact, it even has fewer. 
The only plausible reason why quiche produces substantially more ACK packets than the other stacks is its \texttt{eliciting\_threshold}, i.e., the number of ack-eliciting packets a receiver usually obtains before emitting an ACK packet. The quiche uses a threshold of 1, whereas the others use 2. This also aligns with the observed reply-packet ratios (100\% vs. 50\%).
The QUIC standard states that a receiver should send an ACK after receiving two or more ACK-eliciting packets. 
Both behaviors observed—sending an ACK for every ack-eliciting packet or for every two ack-eliciting packets—remain fully compliant with this specification.
\begin{table}[htbp]
\centering
\caption{Summary of ACK Triggering Behaviors in Three QUIC Implementations}
\renewcommand{\arraystretch}{1.2}
\begin{tabular}{p{1.2cm} p{1.2cm} p{1.6cm} p{1.6cm} p{1.6cm} p{1.6cm}}
\toprule
\textbf{Stack} &
\textbf{Eliciting threshold} &
\textbf{Handshake packets} &
\textbf{Reach max ack delay} &
\textbf{Out-of-order} &
\textbf{IP-ECN}\\
\midrule

quiche &
1 &
\checkmark &
\checkmark &
Not found &
Not found \\
\hline

quinn &
2 &
\checkmark &
\checkmark &
\checkmark &
\checkmark \\
\hline

quic-go &
2 &
\checkmark &
\checkmark &
\checkmark &
\checkmark \\
\bottomrule
\end{tabular}
\label{tab:source-analyze-ack}
\end{table}

\smalltitle{Summary of ACK-Triggering Behaviors.}
As shown in the \cref{tab:source-analyze-ack}, for the vast majority of packets carrying ACK frames, the ultimate triggering event is the reception of a packet, such as the arrival of a handshake packet, arrival of a data packet and reaching the ACK-eliciting threshold, receiving an out-of-order packet, or receiving a packet with IP-ECN markings.
The only exception is the \texttt{max\_ack\_delay} timer.
This observation suggests that if a middlebox has an estimate of the RTT to the receiver, it may be able to identify such correspondences between sent packets and subsequent replies; conversely, if such correspondences can be inferred, they can provide RTT samples.
However, due to RTT estimation uncertainty and limited visibility into receiver-side behaviors, this correspondence is inherently imperfect.
Next, we analyze a strawman solution and show that it cannot maintain correct sent--reply correspondence over time.
We then identify the opportunity provided by traffic locality, which motivates our flowlet-based design in \cref{sec:design}.

\subsection{A Strawman Approach to Matching Reply Packets}

\begin{figure}
    \centering
    \includegraphics[width=0.6\linewidth]{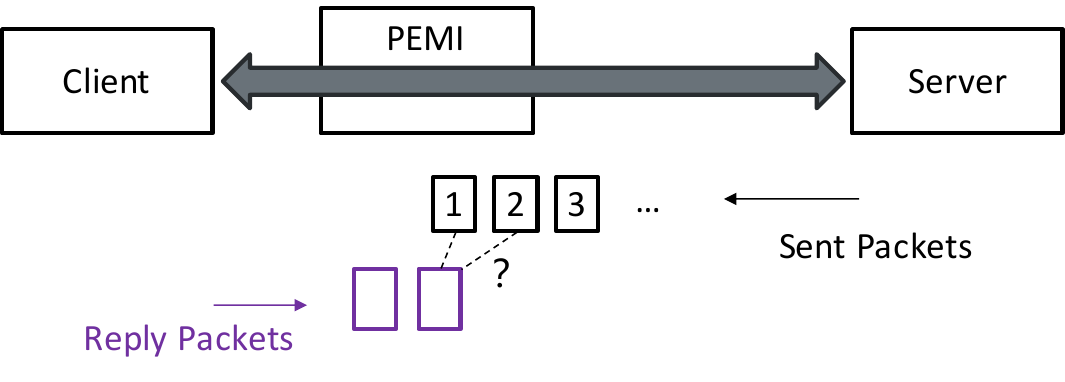}
    \caption{Strawman solution for loss detection.}
    \Description{xxx.}
    \label{fig:strawman_loss_det}
\end{figure}

Traditionally, transport protocols and middleboxes globally match ACKs with sent packets by assigning a unique identifier to each packet.
The strawman solution mimics such global matching.
Specifically, it tries to correspond the data packets with the ACK packets via inference.
As shown in \cref{fig:strawman_loss_det}, the strawman solution treats every reply packet as an ACK and attempts to match it to a sent packet. Since this matching relies on inference without explicit packet numbers, errors are inevitable.
The problem with this strawman solution goes beyond limited accuracy: in the long term, the accuracy will eventually drop to zero.
This is because that \textit{an erroneous match may lead to incorrectness of all subsequent correspondences}.
Specifically, determining which data packet a reply packet corresponds to affects the selection of the starting packet for subsequent reply packets. This could lead to the accumulation of errors. 
It is almost impossible to make continuous correct correspondences.

\subsection{Traffic Locality}
\label{sec:traffic_locality}

Since global loss inference suffers from error accumulation, we seek natural boundaries that can localize inference.
Some application traffic exhibits temporal locality in the form of packet bursts, which provide such inference boundaries.
Specifically, when gaps of these bursts are large enough, a middlebox can easily match the reply packets to a small region of sent packets, and enable further inference within this region.
Consider two consecutive packet bursts separated by an idle gap of length~$t$.
When a reply packet arrives, a middlebox can estimate the timestamp of its corresponding sent packet by subtracting the measured RTT and assign the reply to the temporally closest burst.
Under this nearest-burst assignment, as long as the estimation error (e.g., due to RTT noise) is smaller than $t/2$, the reply packet will be associated with the correct burst.
This tolerance prevents errors from propagating across bursts.

In this paper, we term these bursts \textit{flowlets}~\cite{flare2007sigcomm}.
A flowlet is separated from other flowlets by a large enough time interval called \textit{flowlet timeout}. Flowlets have been extensively investigated in load balancing~\cite{flare2007sigcomm,conga2014sigcomm,nsdi17letitflow}. In these works, the flowlet timeout is set from tens of µs to hundreds of ms, depending on the network conditions and the connection traffic.

Flowlets widely exist in connections, stemming from temporarily limited application data or restrictions imposed by the transport layer.
For instance, in real-time video conferences and cloud gaming, once a frame is completely transferred, it will wait for the generation of the next frame; user interactions may lead to intermittent data transmission of Internet of Things (IoT) applications; and the transfer of new data of Web might occur after the completion of some script execution or rendering.
The congestion control or flow control mechanisms in the transport layer may also result in periods where no packet is allowed to be transmitted.

\noindent\textbf{Intuitive Evidence of Flowlets.}
We examine traffic from deployed applications to identify the presence of
flowlets. 
Specifically, we use the Chrome browser to access websites and watch videos, while capturing QUIC traffic from servers to the client.
QUIC packets are filtered by UDP port~443, and individual QUIC connections are distinguished using five-tuples.

\begin{figure}
    \centering
    \subfloat[Loading \texttt{amazon.com}.]{\includegraphics[width=0.48\linewidth]{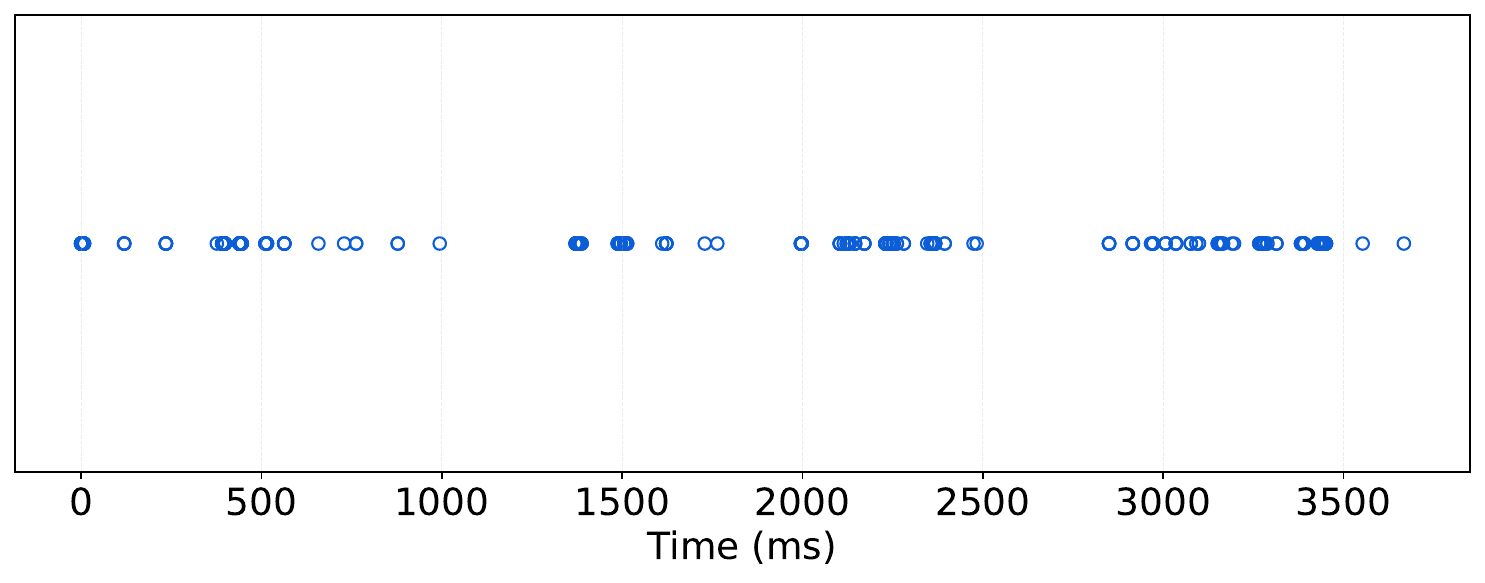}\label{fig:flowlet_web}}
    \hspace{0.02\linewidth}
    \subfloat[Watching a video on \texttt{moq.dev}.]{\includegraphics[width=0.48\linewidth]{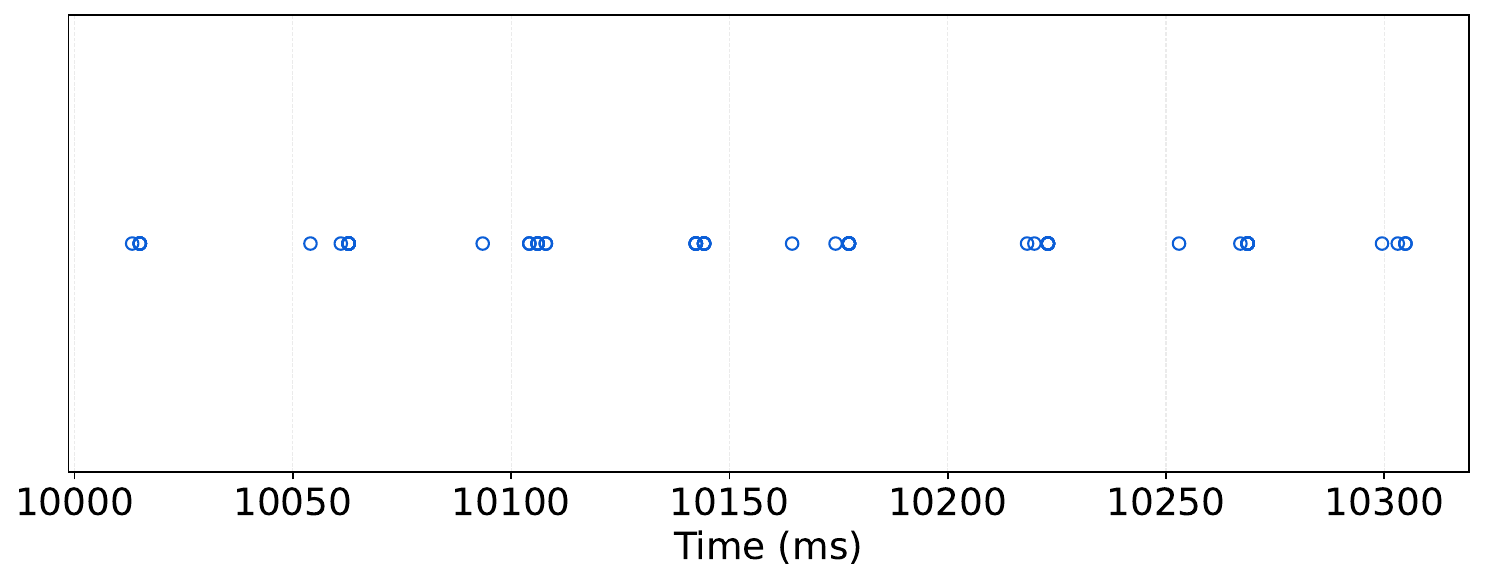}\label{fig:flowlet_moq}}
    \caption{Packet transmission time series of a QUIC connection.}
    \Description{XXX}
    \label{fig:flowlet_examples}
\end{figure}

We first consider webpage loading. \Cref{fig:flowlet_web} shows the packet arrival times of server-to-client packets from a QUIC connection when visiting
\texttt{amazon.com}. 
The transmitted packets clearly form a sequence of small packet
bursts, separated by noticeable idle gaps ranging from several milliseconds to
hundreds of milliseconds.

We next analyze traffic while watching the demo on \texttt{moq.dev}, which is based on the Media over QUIC (MOQ)~\cite{moqgroup}.
We select the connection with the largest data volume, which likely corresponds to the main media stream.
To focus on steady-state behavior, we show a segment of 100 packets starting from 10~seconds of this connection.
As shown in \Cref{fig:flowlet_moq}, packets are transmitted periodically, forming distinct flowlets with an inter-burst interval of approximately 30~ms.

\noindent\textbf{Measuring Flowlet Locality in Top 100 Websites.}
We further investigate the traffic locality of more HTTP/3 web traffic to quantify the prevalence and characteristics of flowlets at scale.
We use the Playwright framework with Chromium to visit the Tranco Top~100 websites~\cite{LePochat2019}, based on the list generated on 03 January 2026.
We filter QUIC traffic by UDP port~443 and distinguish individual QUIC connections using the five-tuple. 
For each connection, we take the direction with larger traffic volume as the sending direction and analyze flowlets in this direction.
Among the 100 websites, we observe QUIC traffic in 18 domains.

\begin{figure}[htpb] 
    \centering
    \subfloat[Avg packet num vs. flowlet gap.]{\includegraphics[width=0.35\linewidth]{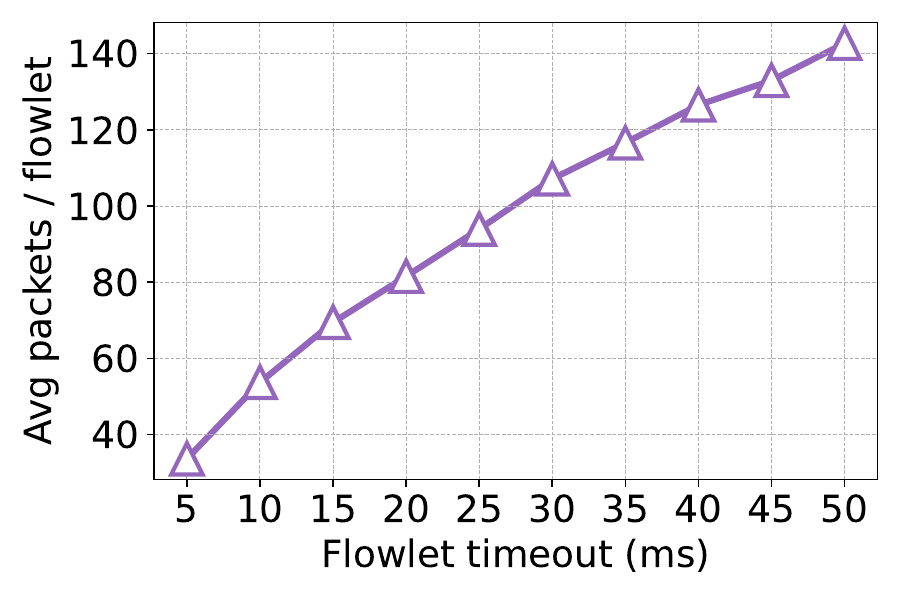}\label{fig:flowlet_pkts}}
    \hspace{0.05\linewidth}
    \subfloat[Avg flowlet duration vs. flowlet gap.]{\includegraphics[width=0.35\linewidth]{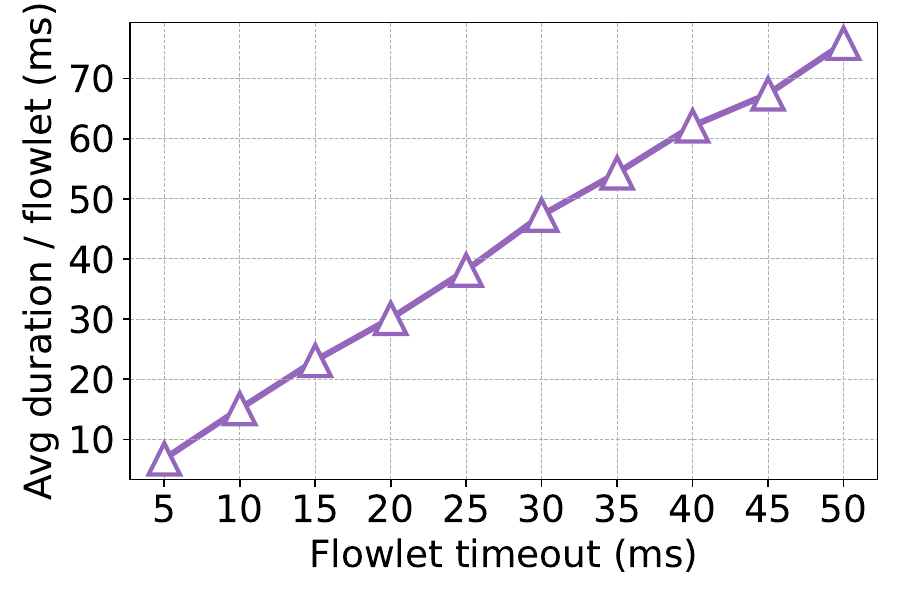}\label{fig:flowlet_duration}}
    \caption{Flowlet measurements from QUIC traffic of Top 100 websites.}
    \Description{xxx.}
    \label{fig:flowlet_vs_timeout}
\end{figure}

The results are shown in \cref{fig:flowlet_vs_timeout}.
The \cref{fig:flowlet_pkts} plots the average number of packets per flowlet under different flowlet timeout values, while the \cref{fig:flowlet_duration} plots the average flowlet duration.
Overall, we observe that clear packet gaps exist and can be used to partition traffic into flowlets.
For example, if the flowlet timeout is set to 20~ms, the average number of packets per flowlet is about 80, and the average flowlet duration is about 30~ms.
Further, both the flowlet duration and the number of packets per flowlet increase approximately linearly with the flowlet timeout.
These results indicate that traffic can be partitioned into flowlets across a wide range of flowlet timeout values, albeit with different flowlet sizes.
Larger gaps make it easier to associate reply packets with the correct flowlet, but also increase the difficulty of inference within a flowlet, such as matching a reply packet to the corresponding sent packet.

Taken together, this section shows that opportunities for transparently identifying correspondences between sent and reply packets do exist in QUIC.
Moreover, by leveraging traffic locality, such correspondence inference can be confined to small regions, which helps prevent error propagation and improves inference accuracy.
However, these opportunities are not always available.
For example, reply packets in QUIC may be triggered by timers or simply be application packets in the reverse direction, in which case there may be no specific sent packet to associate with; some traffic may not exhibit clear locality, and abrupt RTT variations can cause reply packets to be assigned to incorrect flowlets.
These observations form the basis for our subsequent design.

\section{\name Design}
\label{sec:design}

\begin{figure*}[t]
    \centering
    \includegraphics[width=0.95\linewidth]{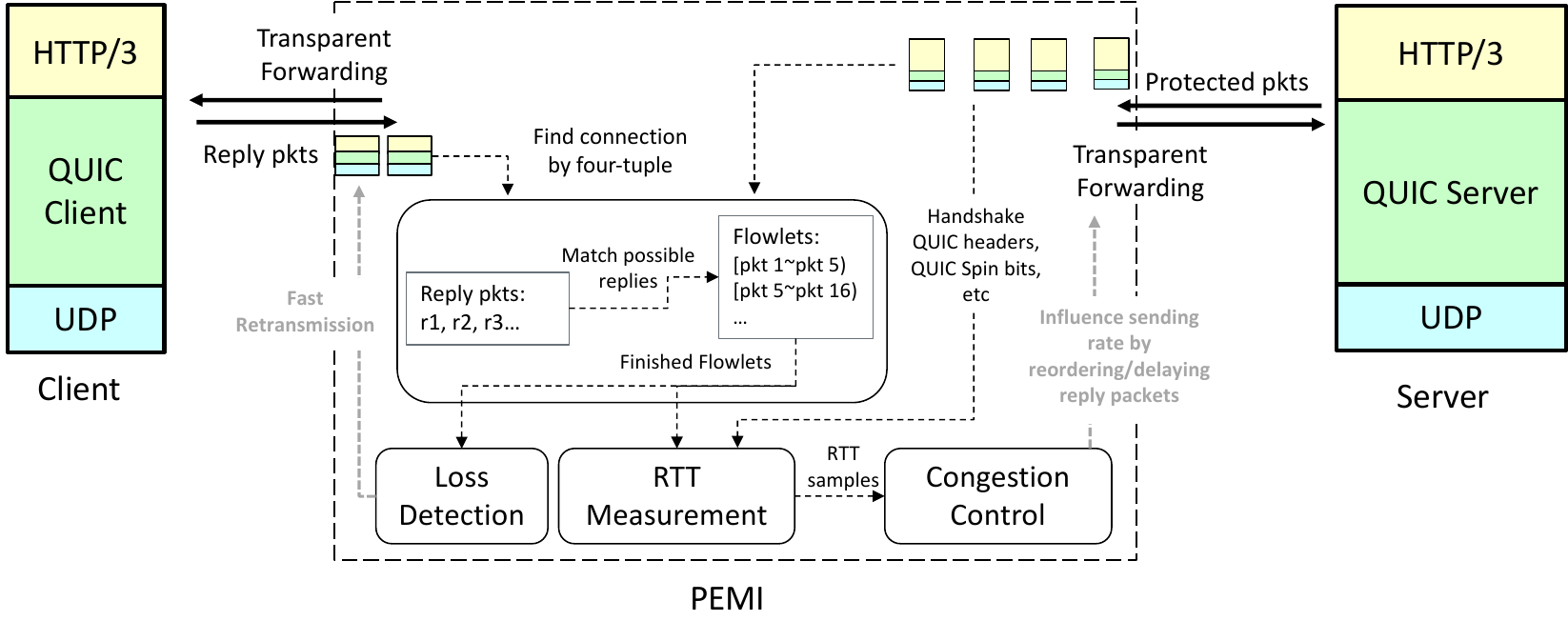}
    \caption{Overview of \name.}
    \Description{XXX}
    \label{fig:overview} 
\end{figure*}

\cref{fig:overview} illustrates the overview of \name. 
\name receives the UDP packets from both sides of the connection, and then forwards them to the other side. It records the flowlets and buffers packets for retransmission.
\name continuously measures the total data volume from both ends. When a \textit{dominant direction}~(\cref{sec:dominant_direction}) of data transmission exists, \name finds reply packets~(\cref{sec:match_packets}) and executes loss detection and fast retransmission~(\cref{sec:loss_recovery}) for sent packets of this direction.
The dominant direction in the \cref{fig:overview} is from the server to the client.
\name continuously measures the RTT from \name to both ends~(\cref{sec:rtt_measurement}) to provides signals for loss detection and congestion control~(\cref{sec:congestion_control}).

\subsection{Dominant Direction}
\label{sec:dominant_direction}

\name only protects packets of dominant direction because the middlebox is unable to reliably determine whether a packet is a data packet or an ACK packet~\cite{rfc9312}. 
Further, QUIC packets may piggyback ACK frames, which can affect the receiver’s ACK packets triggering (\cref{sec:quicobservations}).
If there are many data packets sent from a side, it becomes very difficult to infer whether the packets received by this side have been acknowledged.
\name finds the dominant direction by comparing the total data volume of the two directions.
The volume is counted every end-to-end smoothed RTT.
\name counts total data volume from a side as:
\[
    V = \sum_{\text{i=k}} (B_i-B_\text{min})
\]

Where $B_i$ is the size of the $i$ packet, and $k$th packet is the first one of this turn of counting. We normalize the volume by subtracting the minimum packet size of this connection $B_\text{min}$ because QUIC packets contain considerable header overhead. 
If $2 \cdot V_{\text{client}}  < V_{\text{server}}$, \name protects the packets sent from the server to the client. Conversely, if $V_{\text{client}} > 2 \cdot V_{\text{server}}$, \name protects the packets sent from the client.
Otherwise, it uses the last dominant direction.

\subsection{Matching sent-reply packets}
\label{sec:match_packets}

In \name, we leverage the locality of data transmission to match sent and reply packets within each flowlet, and then measure the RTT and infer potential packet loss events.
This approach helps avoid error accumulation~(see \cref{sec:traffic_locality}). 

\subsubsection{Flowlets recording}
When a protected packet (packet belonging to dominant direction, see~\cref{sec:dominant_direction}) comes, \name records it into a flowlet based on its arrival timestamp and then forwards it. Specifically, if the arrival time of the packet is within the flowlet timeout of the last packet in an existing flowlet, it is added to this flowlet; otherwise, a new flowlet is created. 
The flowlet timeout is set as:
\begin{equation}
    F_{\text{timeout}} = \alpha \cdot sINV \nonumber
\end{equation}

where $sINV$ is the smoothed packet interval of the transmission direction, and $\alpha$ is a constant. We set $\alpha$ as 2 in our experiments in \cref{sec:eval}.

\subsubsection{Identifying Reply Packets for Each Flowlet}
\label{sec:match_to_flowlet}
QUIC fully encrypts ACK information, and the IETF QUIC community explicitly discourages using packet size to distinguish between data packets and ACK packets~\cite{rfc9312}.
\name adopts a conservative approach: takes all packets in the reply direction as potential ACK packets.

For each reply packet, \name primarily relies on a nearest-flowlet association based on packet timestamps and the RTT measured (\cref{sec:rtt_measurement}) between \name and the dominant receiver.
Specifically, upon observing a reply packet arriving at time $t_{\text{reply}}$, \name projects it back to the sender timeline by computing $\tilde{t} = t_{\text{reply}} - \text{RTT}$, and assigns the reply to the flowlet whose time span is temporally closest to $\tilde{t}$.

\name excludes reply packets that arrive significantly earlier or later than any existing flowlet.
Such packets are more likely to be triggered by other causes, such as reverse-direction application data, rather than by sent packets within the current flowlets.
Concretely, if a reply packet’s arrival time $t_{\text{reply}}$ falls outside a global exclusion window $[t_{\text{start}}, t_{\text{end}}]$ defined below, it is discarded and not considered for flowlet matching.

\begin{equation}
    \label{eq:flowlet_reply_find}
    \begin{aligned}
        \Delta_{\text{margin}} &= \beta \cdot F_{\text{timeout}}, \\
        t_{\text{start}} &= t_{\text{flowlets\_begin}} + \text{RTT} - \Delta_{\text{margin}}, \\
        t_{\text{end}} &= t_{\text{flowlets\_end}} + \text{RTT} + \Delta_{\text{margin}}.
    \end{aligned}
\end{equation}

Here, $\Delta_{\text{margin}}$ represents the jitter tolerance used to filter out replies that are unlikely to correspond to any existing flowlet.
$\beta$ is a constant, set to $0.5$ in our experiments.
$t_{\text{flowlets\_begin}}$ and $t_{\text{flowlets\_end}}$ denote the timestamps of the first and last packets among all currently active flowlets, respectively.

\subsubsection{Matching Reply-Sent Packets within a Flowlet}
\label{sec:match_in_flowlet}

\begin{figure}
    \centering
    \includegraphics[width=0.6\linewidth]{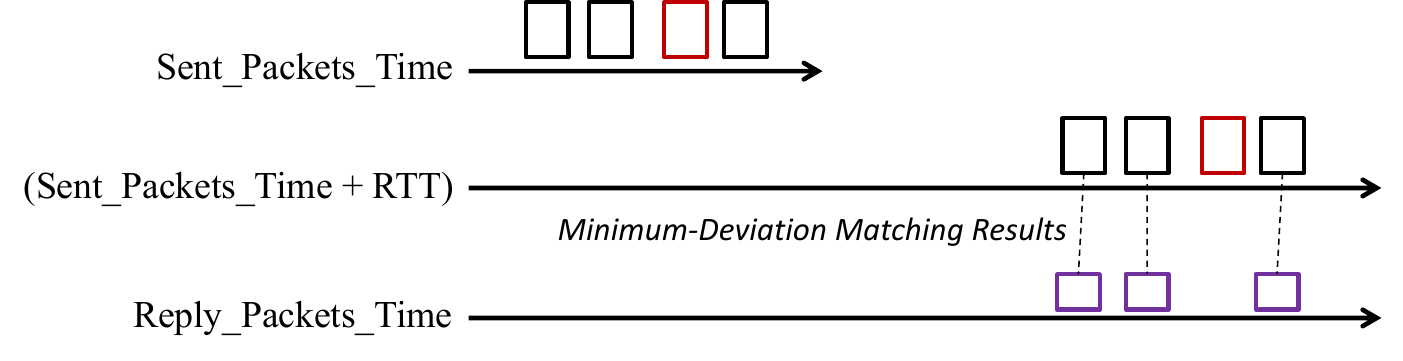}
    \caption{Matching the reply packets and sent packets for a partly replied flowlet.}
    \label{fig:flowlet_loss_det}
\end{figure}

After assigning reply packets to flowlets, \name matches sent packets with reply packets within each flowlet. These matches will be used for packet loss inference and RTT sample computation later.
For each flowlet, the associated reply packets fall into one of the following cases.
(1) No reply packets, and thus no matching;
(2) Equal numbers of sent and reply packets.
\name assumes a one-to-one correspondence and matches packets in timestamp order;
(3) More reply packets than sent packets.
\name conservatively abandons packet-level matching for this flowlet, as it cannot identify which reply packets are not triggered by sent packets;
(4) Fewer reply packets than sent packets.
\name computes a matching between sent and reply packets using a dynamic programming (DP) algorithm.
Specifically, as shown in Figure~\ref{fig:flowlet_loss_det}, \name attempts to find the correspondence that minimizes the total deviation between each matched pair's timing gap and the measured RTT:

\[
\min_{\mathcal{M}}
\sum_{(i,j)\in\mathcal{M}}
\bigl|\,(t^{\text{reply}}_{j}-t^{\text{sent}}_{i}) - \widehat{\text{RTT}}\,\bigr|,
\]
where $\mathcal{M}$ denotes a matching between sent and reply packets, $t^{\text{sent}}_i$ and $t^{\text{reply}}_j$ are the timestamps of the $i$-th sent packet and the $j$-th reply packet observed at the middlebox, respectively, and $\widehat{\text{RTT}}$ is the estimated RTT for the matching of this flowlet.
This method assumes that packets within the same flowlet, which are typically distributed in milliseconds or tens of milliseconds, experience similar RTT.

To compensate for potential drift between the RTT experienced by the current flowlet and the previously measured RTT samples, \name tries to compute a fresh RTT estimate for the matching as follows:
\[
\widehat{\text{RTT}} =
\begin{cases}
t_{\mathrm{last\_reply}} - t_{\mathrm{last\_sent}},
& \text{if } W_{\text{reply}} > 0.8\, W_{\text{sent}}, \\[6pt]
\text{srtt},
& \text{otherwise}.
\end{cases}
\]

When the reply packets within a flowlet appear to align well with the sent packets---simply judged by whether their temporal widths($W_{\text{reply}}$ and $W_{\text{sent}}$) are similar---\name estimates the RTT using the timestamp gap between the last reply packet and the last sent packet.
Otherwise, \name uses the existing measured smoothed RTT (srtt).

\subsection{Loss Detection and Recovery}
\label{sec:loss_recovery}

\subsubsection{Loss detection}
\label{sec:flowlet_loss_det}

Packet loss is inferred within each flowlet.
There are three conditions: 
(1) if the reply packets number is equal to or more than the packets in the flowlet, \name considers all packets in the flowlet have been acknowledged; 
(2) for the flowlet with no reply, \name infers that all packets are lost;
(3) if there are some replies but fewer than the packets sent, \name infers loss among sent packets without a matched reply after the DP-based matching in \cref{sec:match_in_flowlet}.

For partly replied flowlets (condition (3)), \name considers two cases in which a sent packet should not be inferred as lost. This design is based on observations from the QUIC standard and implementations~(\cref{sec:quicobservations}).

\begin{figure}
    \centering
    \includegraphics[width=0.8\linewidth]{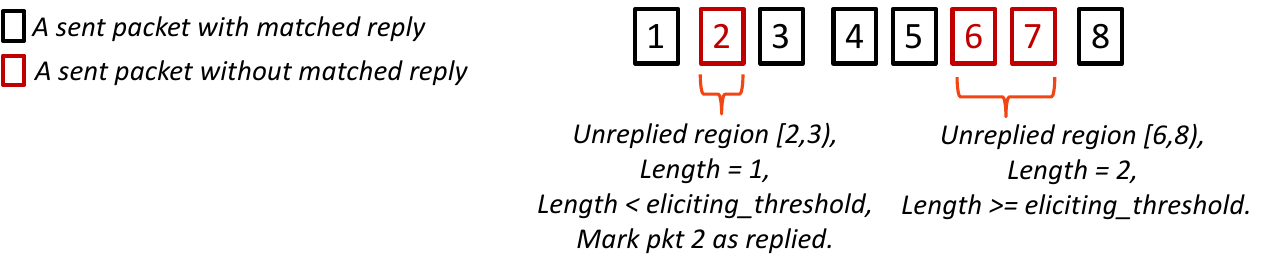}
    \caption{Distinguishing ACK suppression caused by the eliciting threshold. In this example, the \texttt{eliciting\_threshold} is 2.}
    \label{fig:design_2pointer}
\end{figure}

(1) QUIC allows a receiver to delay sending ACK frames until it has observed a certain number of ACK-eliciting packets, as controlled by the \texttt{eliciting\_threshold}.
The eliciting\_threshold is estimated from the ratio of sent to reply packets of the connection~(\cref{sec:impl}). 
When this threshold is greater than one, some received packets may appear unreplied. 
\name applies a post-processing step based on two-pointer scanning.
As shown in the \cref{fig:design_2pointer}, \name linearly scan sent packets of a flowlet using two pointers.
The left pointer identifies the beginning of a maximal consecutive unreplied region, and the right pointer advances until either a matched reply is encountered or the end of the sequence is reached.
For each unreplied region $\left[l, r\right)$, if it is followed by a valid reply and its length is smaller than the \texttt{eliciting\_threshold}, \name interprets this region as the result of ACK aggregation rather than packet loss. 
All packets in this region are therefore marked as logically replied. 
In contrast, longer unreplied regions are preserved as potential loss candidates.

(2) QUIC allows ACKs to be sent after batched processing of multiple received packets.
When the interval between several packets in a flowlet is less than \texttt{close\_interval}, \name considers it likely that they could be ACKed in the same reply packet, and ignores the inferred packet losses for each one of them.
We set \texttt{close\_interval} as 0.1ms in our experiments in \cref{sec:eval}.
This is a conservative strategy to avoid unnecessary retransmissions. Consequently, it may inevitably miss some losses, for example when a lost packet is adjacent to packets that are not lost.

\smalltitle{Time complexity.} 
For a partly replied flowlet with $n$ sent packets and $m$ reply packets, the algorithms to infer lost packets have a time complexity of $O(mn)$.
Specifically, the DP algorithm for matching has a time complexity of $O(mn)$, while the linearly scanning step to handle delayed ACKs has a time complexity of $O(n)$. 
For each sent packet, the computational cost is therefore $O(m)$. Since we bound the number of packets within each flowlet (see \cref{sec:best_effort}), this computation can be regarded as a constant per-packet overhead.

\subsubsection{Fast Retransmission}
A flowlet is considered finished once incoming reply packets are no longer assigned to it.
Specifically, \name associates each reply packet with the temporally closest flowlet based on projected send times~(\cref{sec:match_to_flowlet}).
When a reply packet is assigned to a subsequent flowlet, the earlier flowlet is labeled as finished, because any later reply packets (with later timestamps) will no longer be considered as replies for the finished flowlet.
\name buffers all packets of flowlets that are not yet finished. When a packet is inferred to be lost, \name will retransmit it immediately.
\name cannot ensure that the lost packets it infers are correct, nor has a method to confirm the success of retransmissions. 
Therefore, it does not perform repeated retransmissions to avoid excessive bandwidth consumption of wrong retransmissions and risk of amplification attacks.
Specifically, \name records the retransmitted packets in flowlets (as the receiver may reply to them), but avoids retransmitting them if they are inferred to be lost again.

\subsection{RTT Measurement}
\label{sec:rtt_measurement}

Measuring the \name-to-end RTT is crucial for the loss detection~(\cref{sec:match_packets} and \cref{sec:loss_recovery}) and congestion control~(\cref{sec:congestion_control}) in \name.
End-to-end protocols measure RTT samples by the duration between a packet sending and its ACK arriving.
However, \name cannot directly measure the RTT in this way, since the packet number and ACK of QUIC are encrypted.

\subsubsection{Initial RTT}
At the beginning of a connection, QUIC uses packets with long headers for the handshake. The middleboxes can measure the RTT to each end by these packets~\cite{rfc9312}. \name uses RTT measured during handshake as the initial RTT for the connection.
The details of the QUIC handshake packet tracking are described in~\cref{sec:impl}.

\subsubsection{Continuous RTT Measurement}
\label{sec:rtt_measure}

To help the RTT measurement on middleboxes, some QUIC connections deploy the spin bit~\cite{rfc9000}. Unfortunately, as a recent study~\cite{spin2023imc} shows, only \textasciitilde 10\% of QUIC-enabled domains support the spin bit. Additionally, the QUIC standard specifies that the spin bit must be disabled by part of connections~\cite{rfc9000}.
Moreover, the spin bit has an inherent limitation: it provides only one sample per RTT, which may not capture RTT variations effectively.
There are also some out-of-band methods to measure RTT, such as crafting fake QUIC requests to the server side or \texttt{ping} the host.
However, crafting requests frequently will impose an additional burden on the middlebox and endpoints.
Therefore, \name primarily relies on a flowlet-based mechanism for RTT measurement, while the spin bit and synthetic requests are used only for periodic calibration~(\cref{sec:rtt_reset}).

To measure the RTT continuously, \name uses RTT samples from: 
(1) the exactly replied flowlets, by calculating an RTT sample for each pair of sent-reply packets; and
(2) the partly replied flowlets, using the optimal matching found by the DP algorithm in \cref{sec:match_in_flowlet} to derive RTT samples from each matched sent-reply packet pair.
Flowlets without reply packets cannot provide RTT samples.
We also discard flowlets where the number of reply packets exceeds the number of sent packets, as \name cannot definitively know which of these are not acknowledging the sent packets.
Note that the RTT measurement is \textit{best-effort}, because there is no guarantee that the matching between data packets and reply packets is entirely correct.

\subsubsection{RTT Calibration}
\label{sec:rtt_reset}

Flowlet-based RTT measurement relies on correct reply-to-sent mapping for flowlets.
This approach can track RTT variations up to half a flowlet interval (e.g., about 15~ms for 30~FPS video streams) across consecutive flowlets.
However, abrupt RTT variations that exceed this tracking capability cannot be captured, causing the RTT estimation to fail.

\name could leverage best-effort signals (e.g., spin bits) for periodic calibration to keep the RTT measurement accurate even in the presence of large RTT variations.
Middleboxes can leverage three types of information for RTT measurement:
(1) Spin bits~\cite{rfc9000}. When enabled in QUIC connections, spin bits provide one RTT sample per round-trip;
(2) Large enough intervals. The middleboxes could measure the RTT from send-reply pairs after idle periods. These could be caused by application data sending pauses or upstream congestion;
(3) Actively probe the endpoints and infer RTTs from the responses, using protocols such as ICMP, QUIC, or application-layer protocols (e.g., HTTP).
All these methods are best-effort and cannot guarantee applicable RTT calibration across all connections.

\cref{tab:rtt_reset_methods} compares the limitations of these approaches.
For active probing, we recommend using ICMP in practice, as it imposes the least computational overhead on the endpoint.
The response rate to ICMP is generally higher than that of other protocols in the Internet~\cite{scan2018}.
If the protected endpoint is behind a NAT, ICMP responses can still be obtained as long as the middlebox is within the same LAN, such as when it serves as the access point for the endpoint.
The additional bandwidth consumption introduced by reasonable probing frequencies is negligible. 
For example, for flowlets consisting of approximately 10 packets each, sending one ping request for every 5 flowlets results in only about 2\% additional packets and consumes roughly 0.1\% extra bandwidth (assuming each data packet is 1000 bytes). Since RTT is measured between the middlebox and the protected endpoint, this bandwidth overhead only affects the sub-link between them.

\begin{table}[t]
    \centering
    \caption{Comparison of Middlebox-Available RTT Calibration Methods}
    \begin{tabular}{l|l|l}
        \toprule
        \multicolumn{2}{l|}{\textbf{Methods}} & \textbf{Limitations} \\
        \midrule
        \multicolumn{2}{l|}{QUIC spin bit~\cite{rfc9000}}  & Only one sample per RTT; Limited deployment in the wild~\cite{spin2023imc}. \\ \hline
        \multicolumn{2}{l|}{Packets after idle periods  }       & Requires large packet intervals.       \\ \hline
        \multirow{2}{*}{Active probing} & ICMP       & Potential non-response~\cite{scan2018}; Only network layer RTT.       \\ \cline{2-3}
                & QUIC       & Only available for servers; Endpoint computation overhead.       \\
        \bottomrule
    \end{tabular}
    \label{tab:rtt_reset_methods}
\end{table}

\subsection{Black-Box Congestion Control}
\label{sec:congestion_control}
\name employs a congestion control due to the following reasons:
(1) Fairness with flows. The fast retransmission of \name may lead to fewer losses observed at the QUIC endpoint. 
By using congestion control, \name could control the aggressiveness of protected connections, keeping the fairness with other flows.
(2) Congestion avoidance. \name executes fast retransmissions transparently to the endpoint. If packet loss events are due to congestion dropping, unlimited retransmissions are harmful to the network. Additionally, the end hosts might retransmit the same packets as the \name, because it may still detect a packet as lost due to reasons such as out-of-order ACKs. This will lead to a waste of bandwidth and make the congestion worse.
Therefore, the \name needs a mechanism to detect congestion and reduce retransmissions when the link is already congested.

However, implementing a congestion control for \name is challenging. Existing congestion control algorithms are designed for end-to-end protocols, which are able to see the congestion signals, usually on ACK arrival. Furthermore, the endpoints can enforce the congestion window (CWND) by limiting the data sending. \name achieves congestion control in a completely transparent way. It utilizes the inferred RTT samples as the congestion signals and employs a feedback-based method to enforce the CWND.

\subsubsection{Delay-based Congestion Control Algorithm.}

We develop congestion control in \name based on Copa~\cite{copa2018nsdi}, a congestion control algorithm that relies on RTT-derived congestion signals measurable by \name~(\cref{sec:rtt_measurement}).
Copa is an ACK-driven algorithm that calculates a new RTT sample and adjusts CWND every time an ACK arrives. It decides whether to speed up or slow down based on RTT samples and CWND. When deciding to speed up, Copa increases the CWND by: 
\[
\text{CWND} = \text{CWND} + \frac{\nu}{(\delta \cdot \text{CWND})}
\]
where $\nu$ is a dynamical parameter and $\delta$ is the parameter to control the aggressiveness of the algorithm. When deciding to slow down, Copa decreases the CWND by:
\[
\text{CWND} = \text{CWND} - \frac{\nu}{(\delta \cdot \text{CWND})}
\]
Over one RTT, the change in CWND is $\approx \nu / \delta$ packets.

However, \name cannot adjust the CWND at per-ACK granularity. 
To achieve the similar convergence speed, \name adjusts the CWND every time the RTT is updated, and the change in CWND is:
\[
     \frac{min(\Delta t, srtt)}{srtt} \cdot \frac{\nu}{\delta}
\]
where $srtt$ is the measured smoothed RTT, and $\Delta t$ is the time interval from the last CWND adjustment. This modification makes \name keep the similar convergence speed as Copa, changing the CWND by $\approx \nu / \delta$ packets over one RTT. For a single time of update, \name adjusts the CWND by at most $\nu / \delta$ packets (when $\Delta t >= srtt$).

\subsubsection{Enforce the CWND}
\name does not directly use the computed CWND. Because it cannot accurately track the remaining CWND due to the lack of plaintext ACKs.
Alternatively, for each connection, \name controls its packet sending rate, which is the sum of the forwarding rate and the retransmission rate.
\name computes the target sending rate by:
\begin{equation}
    \text{rate}_\text{target} = \frac{\text{CWND}}{\text{RTTmin}}
\end{equation}
where $\text{RTTmin}$ is the minimum RTT in a long period of time.

\name enforces the target rate by a feedback-based control loop. 
It measures the actual sending rate by the number of packets sent in a period of time.
When the actual rate is greater than the target rate, \name stops the fast retransmissions. If the remaining rate is still greater than the target rate, \name will \textit{reorder or delay the reply packets} to make the overspeed sender slow down. 
QUIC detects packet loss using both packet-based and time-based thresholds, denoted as \texttt{kPacketThreshold} and \texttt{kTimeThreshold}~\cite{rfc9002}, which are recommended to be set to 3 and 9/8, respectively.
Once a later packet is acknowledged, an earlier packet $p$ is declared lost if either the acknowledged packet was sent at least three packets after $p$, or more than 9/8 RTT has elapsed since $p$ was sent.
Accordingly, \name may trigger both packet-threshold-based and time-threshold-based loss detection in the sender by introducing sufficient packet reordering or forwarding delays.
For senders that have not slowed down after one end-to-end RTT, \name doubles the parameters of the reordering and delaying thresholds for it.

We choose the reply packets delaying and reordering as feedback because (1) compared to manipulating sent packets, changes in the reply packets can reach the sender and influence the end-to-end sending rate more quickly; (2) delaying and reordering, instead of dropping, are less harmful to data transmission. For example, if the reply packets contain data, reordering will usually cause them to arrive slightly later, without the need to wait for the retransmission from the peer.

\subsection{\name is Best-Effort}
\label{sec:best_effort}
The \name is a best-effort solution. When effective inferring is not possible, or it may negatively impact the link, \name will give up the enhancements. Specifically, \name stops enhancements in the following situations:
\begin{itemize}[leftmargin=*] 
    \item Inability to effectively distinguish flowlets. This happens when the packet intervals are too uniform, which may be a result of good network conditions and the application continuously having enough data.
    Fortunately, connections with good network conditions do not urgently require protection from \name.
    Specifically, when the number of packets in a flowlet exceeds $N$, \name abandons the protection for it. We set $N$ as 100 in our experiments in \cref{sec:eval}.
    \item Excessively high detected packet loss rate. 
    This situation may be because of extremely poor network conditions, or the end points implement a very low ACK frequency.
    This may also be caused by wrong RTT estimation, where numerous replied packets are falsely classified as lost.
    \name sets an upper limit on bandwidth utilization to avoid excessive network bandwidth consumption and negative impacts. In our experiments, \name temporally cancels the retransmission when the number of retransmitted packets is greater than 10\% of the number of forwarded packets.
\end{itemize}
\section{Implementation Overview}

We implement a prototype of \name in Rust with approximately 3K lines of code.
Because the QUIC stacks used in our evaluation do not support the spin bit, the current implementation performs RTT calibration only using ICMP probes.
\name collects RTT samples from ICMP probing every end-to-end RTT, same as the sampling frequency provided by the spin bit.
Further implementation details are provided in \cref{sec:impl}.

\section{Evaluation}
\label{sec:eval}
We extensively evaluate \name to answer the following questions: 

\begin{enumerate}[leftmargin=*]
    \item Could \name substantially improve the performance of QUIC?~(\cref{sec:eval_end2end} and \cref{sec:eval:mahimahi}) 
    \item Will the optimization of \name cause QUIC to become overly aggressive under packet loss, thereby deviating from TCP-friendly behavior?~(\cref{sec:eval_fairness})
    \item How many computational resources does \name require, and does it exceed the computational capabilities of middleboxes used in the TCP era?~(\cref{sec:eval_cpu})
\end{enumerate}

To answer these questions, we evaluate \name under two applications with distinct traffic characteristics: download flows, which involve transferring a single bulk object, and RTC frames, which consist of multiple time-sensitive objects.
For download workloads, the application used (nginx and curl) supports both QUIC and TCP, enabling a direct comparison between transparent enhancements for QUIC and TCP.
For RTC workloads, we implement dummy servers and clients based on multiple popular QUIC stacks to demonstrate that \name is effective for different stacks.
All evaluated TCP and QUIC endpoints use CUBIC congestion control.
We conduct the tests mainly in the mininet emulation environment, while also evaluating the RTC application under a trace-driven environment.
The detailed experimental setup is described in \cref{sec:eval:setup}.

\subsection{Setup}
\label{sec:eval:setup}

\begin{table}[tbp]
    \centering
    \caption{Mininet Link Setup of \cref{sec:eval}.}
    \begin{tabular}{l|p{3.9cm}|p{3.9cm}}
        \toprule
        \textbf{Section} & \textbf{Link1(RTT,loss,BW)} & \textbf{Link2(RTT,loss,BW)} \\
        \midrule
        \ref{sec:eval_fct_filesize} & 2ms, [0, 1]\%, 100Mbps & 50ms, 0\%, 10Mbps \\
        \hline
        \ref{sec:eval_delay_rtc},\ref{sec:eval_cpu} & 2ms, 1\%, 100Mbps & 50ms, 0\%, 10Mbps \\
        \hline
        \ref{sec:eval_fairness} & 2ms, [0-5]\%, 100Mbps & 50ms, 0\%, 10Mbps \\
        \hline
        \ref{sec:eval:mahimahi} & 2ms, GE Model, 100Mbps * & 100ms, 0\%, 10Mbps \\
        \bottomrule
    \end{tabular}
    \par\smallskip
    \noindent\footnotesize{*: Except the tc links, The \cref{sec:eval:mahimahi} creates a CellReplay-based trace replay link.}
    \label{tab:eval_setup}
\end{table}

\subsubsection{Application1: HTTP/3 download flows.}
We evaluate the performance of \name in enhancing the goodput of QUIC during file transfers with different sizes.
We use the nginx~\cite{quiche_nginx} as the HTTP server and curl~\cite{curlquiche} as the client, both with HTTP/3 supported by quiche~\cite{quiche}.
Since the application used in this experiment supports both TCP and QUIC, we conduct an apples-to-apples comparison between \name for QUIC and PEP for TCP.
We select PEPSAL~\cite{pepsal2006} as the PEP implementation to improve the performance of TCP. 
PEPSAL splits the end-to-end TCP connection by the SYN packets during the handshake process. It creates split connections and communicates with real endpoints.
For the TCP baseline, we use curl in HTTP/2 mode.

\subsubsection{Application2: RTC frames.}
\label{sec:app_rtc}
We implement QUIC-based dummy servers and clients to send media frames at 30fps, 3000kbps.
To validate whether \name is effective across multiple QUIC stacks, we implement it on the three popular (most starred on GitHub) open-source QUIC stacks: quic-go, quiche, and quinn.
The server sends a frame every 33ms. We measure two kinds of statistics:
\begin{itemize}[leftmargin=*]
    \item Frame-level jitter: statistical variance of the RTP frame interarrival time. 
    \item Frame delay: time between the frame created at the server and completely received at the client.
\end{itemize}

The definition of these two metrics is the same as~\cite{imc2022zoom}.

\subsubsection{Mininet-based Emulation Environment.}
We emulate scenarios that are well-suit to be optimized by PEP/\name, such as \cref{fig:pep_scenario}. The enhancement solution is deployed in a middlebox, and there is a lossy link between it and the protected end.
We conduct our evaluation environment by Mininet~\cite{mininet}. 
Mininet enables flexible modification and validation of diverse network metrics (e.g., packet loss rate), while supporting real-world application deployment (e.g., Nginx web server) and protocol implementations.
All emulation experiments are run on an x86-64 Linux machine equipped with an 8-core 3.60GHz CPU and 32GB memory.

\begin{figure}
    \centering
    \includegraphics[width=0.66\linewidth]{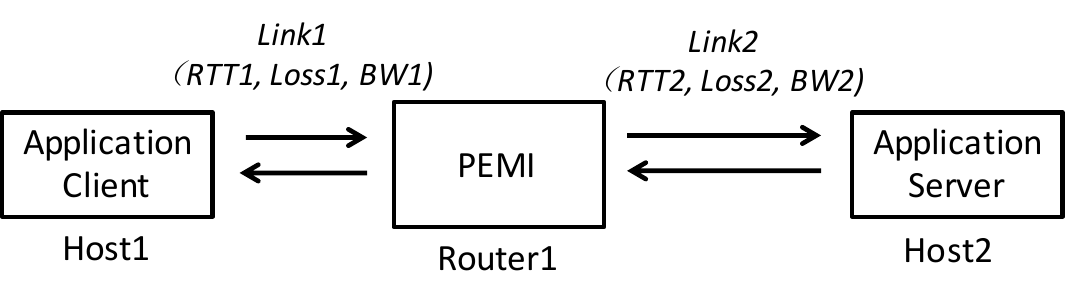}
    \caption{Mininet Emulation Environment.}
    \label{fig:eval_setup}
\end{figure}

As shown in \cref{fig:eval_setup}, the environment consists of three hosts and two links. The Host1 runs the application server; the Host2 executes the application client; and the \name/PEP implementation exists on Router1. 
The links are characterized by specific network attributes such as Round-Trip Time (RTT), random packet Loss rate (Loss), and Bandwidth (BW).
These link parameters are configured using the Linux tc utility.
The connectivity between the client and server crosses two links, Link1 (Host1-Router1) and Link2 (Router1-Host2).
Setup of Links in the following evaluations is shown in \cref{tab:eval_setup}.

\subsubsection{Emulation with Real-Network Traces}
We further evaluate \name under highly dynamic network conditions using real-world traces.
This evaluation is enabled by recent advances in network emulation, namely CellReplay~\cite{sentosa2025cellreplay}.
CellReplay is a Mahimahi-based replay tool that improves the emulation of dynamic delays compared to vanilla Mahimahi. 
It integrates seamlessly with our Mininet testbed: when launching the client on \texttt{HOST1} inside a Mahimahi shell, a dynamic Mahimahi link is inserted between the client and the egress of \texttt{HOST1}.
CellReplay~\cite{sentosa2025cellreplay} provides real cellular network traces.
We use traces collected in driving scenarios, which induce significant network dynamics.
Since CellReplay does not include packet-loss traces, we inject packet loss using a two-state Markov chain model—the Gilbert–Elliott (GE) model~\cite{gemodel,gemodel2,gemodelInternet,gemodelvideo}—implemented in \texttt{tc}.
We emulate loss patterns observed in cellular driving scenarios, characterized by approximately periodic burst losses. Such bursts are likely caused by base station handovers and occur on average every few seconds, with durations on the order of tens of milliseconds~\cite{liu2024m2ho}.
To emulate this behavior, we configure the GE model to transition into a burst-loss bad state (100\% loss rate) approximately every 5~s and to return to the good state (0\% loss rate) after about 50~ms.
Since the GE model in \texttt{tc} only supports per-packet state transitions, we focus on RTC experiments where the goodput remains relatively stable.
Specifically, for 3000~kbps RTC applications, the GE model is configured with a good-to-bad transition probability of 0.069\% and a bad-to-good transition probability of 6.93\%.
The results are presented in \cref{sec:eval:mahimahi}.

\begin{figure}[hbpt]
    \subfloat[0\% random loss.]{\includegraphics[width=0.55\linewidth]{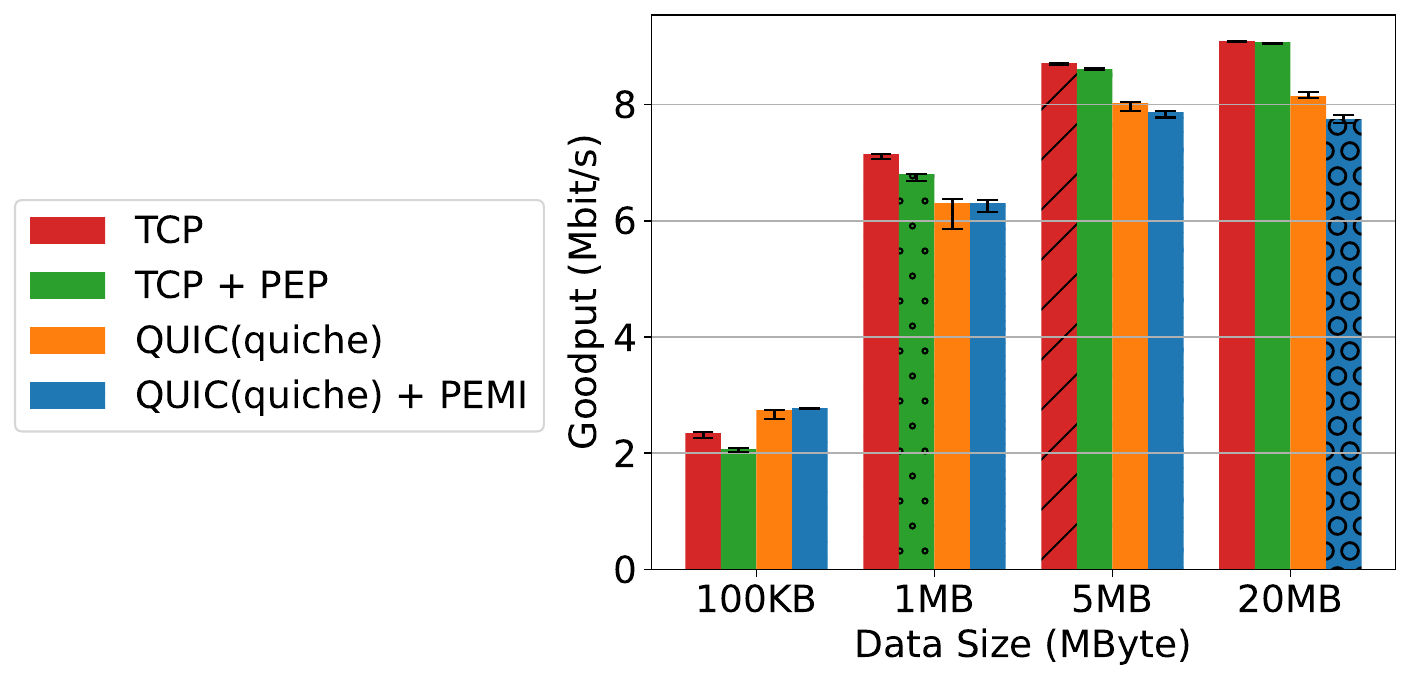}\label{fig:fct_filesize_loss0}}
    \subfloat[1\% random loss.]{\includegraphics[width=0.33\linewidth]{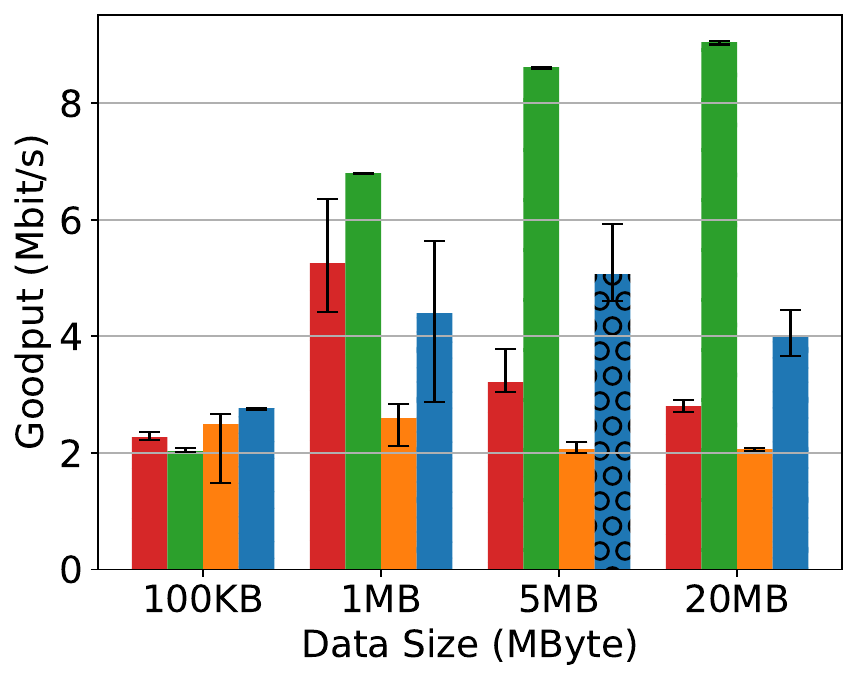}\label{fig:fct_filesize_loss1}}
    \caption{Goodput during file transfers with different sizes. Median of 10 trials. Error bars are 25\% and 75\%.}
    \Description{xxx.}
    \label{fig:fct_filesize}
\end{figure}

\subsection{End-to-End Performance}
\label{sec:eval_end2end}

\subsubsection{Goodput of file transfers.}
\label{sec:eval_fct_filesize}
We measure the performance of \name in enhancing QUIC for the transmission of different flow sizes.
We show the results under 0\% (\cref{fig:fct_filesize_loss0}) and 1\% (\cref{fig:fct_filesize_loss1}) random loss.
As \cref{fig:fct_filesize_loss0} shows, when there is no random loss, both QUIC and TCP exhibited excellent performance. In this scenario, the impact of PEP and \name on their respective enhanced protocols was minimal.

However, as shown in the \cref{fig:fct_filesize_loss1}, when running on links with a little loss (1\%), both QUIC and TCP experienced serious performance degradation. For QUIC connections enhanced by \name, the performance is significantly improved. Specifically, in the transmission of a 5MB file, \name achieves 2.5$\times$ the goodput compared to QUIC. Similarly, PEP increases the goodput of TCP to 3.2$\times$.
This highlights the effectiveness of \name in mitigating the adverse effects of packet loss on network performance.

\subsubsection{Delays of RTC frames}
\label{sec:eval_delay_rtc}

\begin{figure}[hbpt]
    \begin{minipage}[b]{0.65\linewidth}
        \subfloat[Frame-level jitter.]{\includegraphics[width=0.49\linewidth]{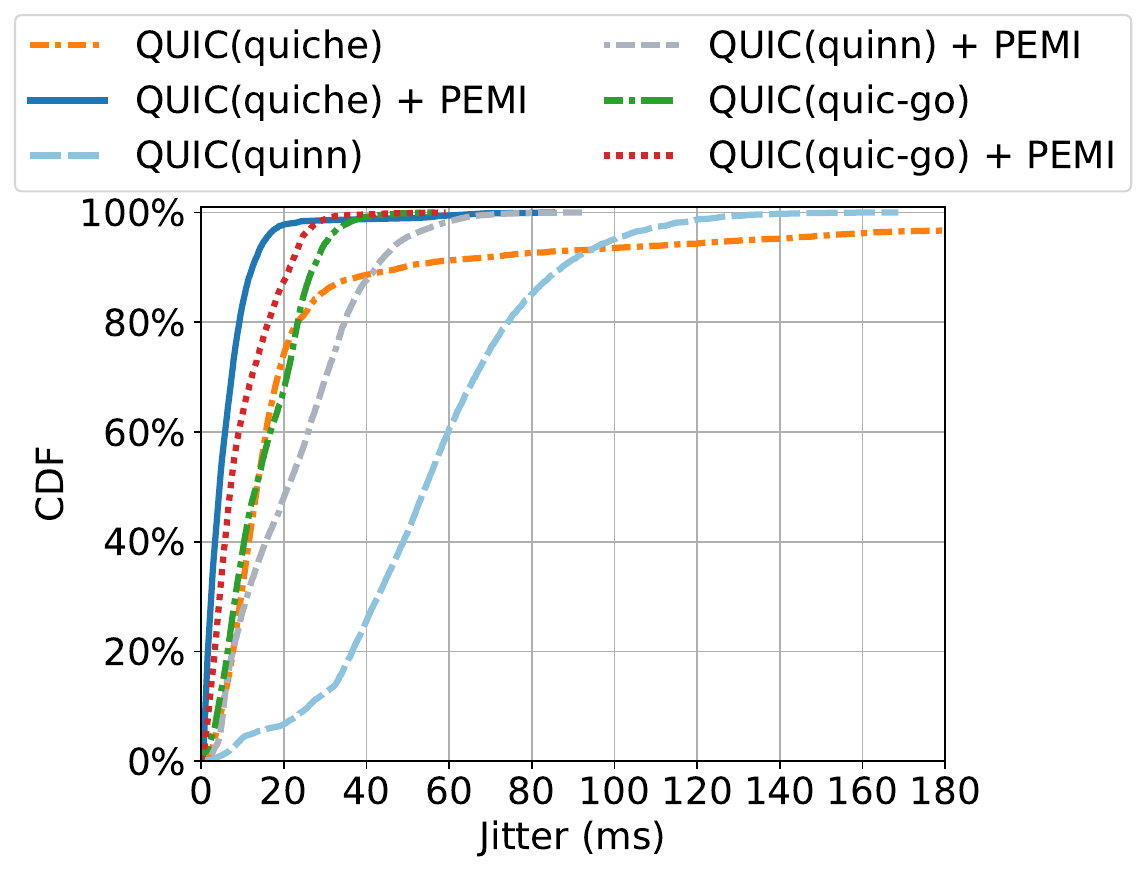}\label{fig:rtc_delay_jitter}}
        \subfloat[Frame delay.]{\includegraphics[width=0.49\linewidth]{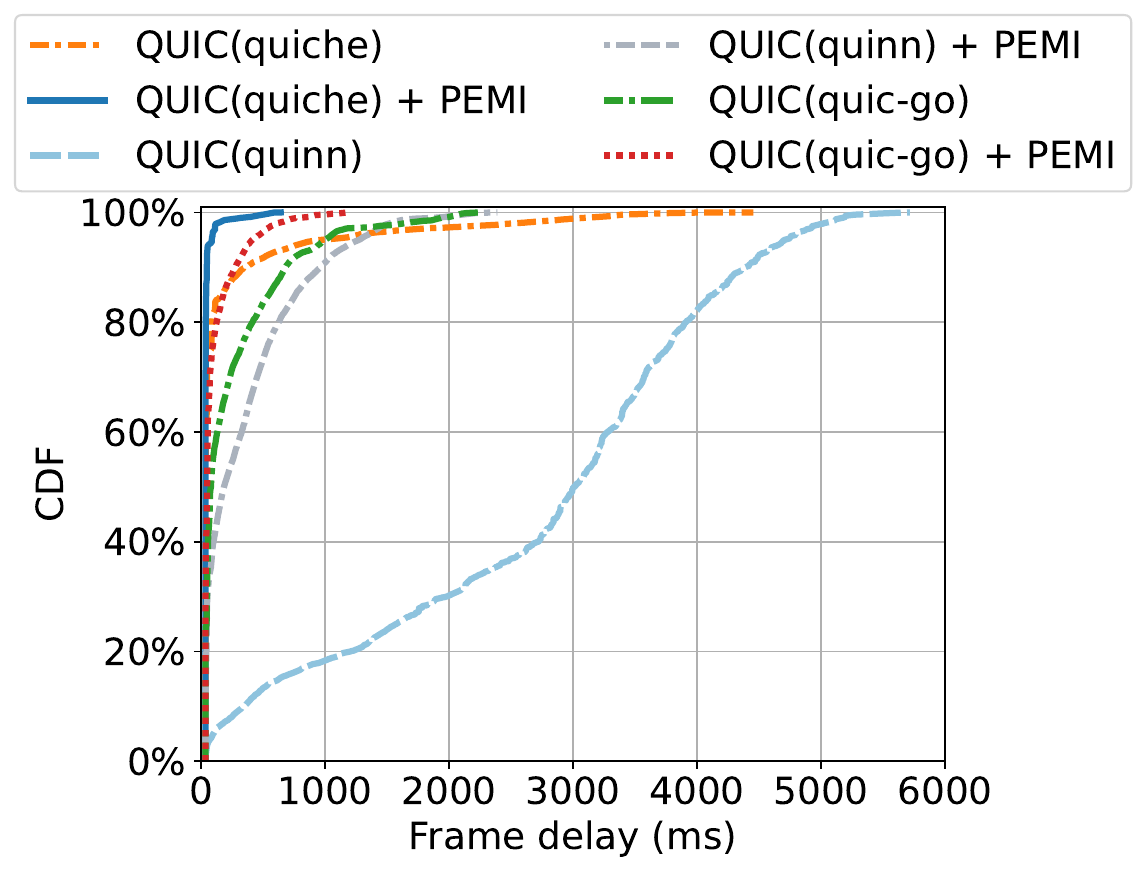}\label{fig:rtc_delay_frame}}
        \caption{Performance of RTC frames transmission.}
        \Description{xxx.}
        \label{fig:rtc_delay}
    \end{minipage}
    \begin{minipage}[b]{0.34\linewidth}
    \centering
    \includegraphics[width=0.9\linewidth]{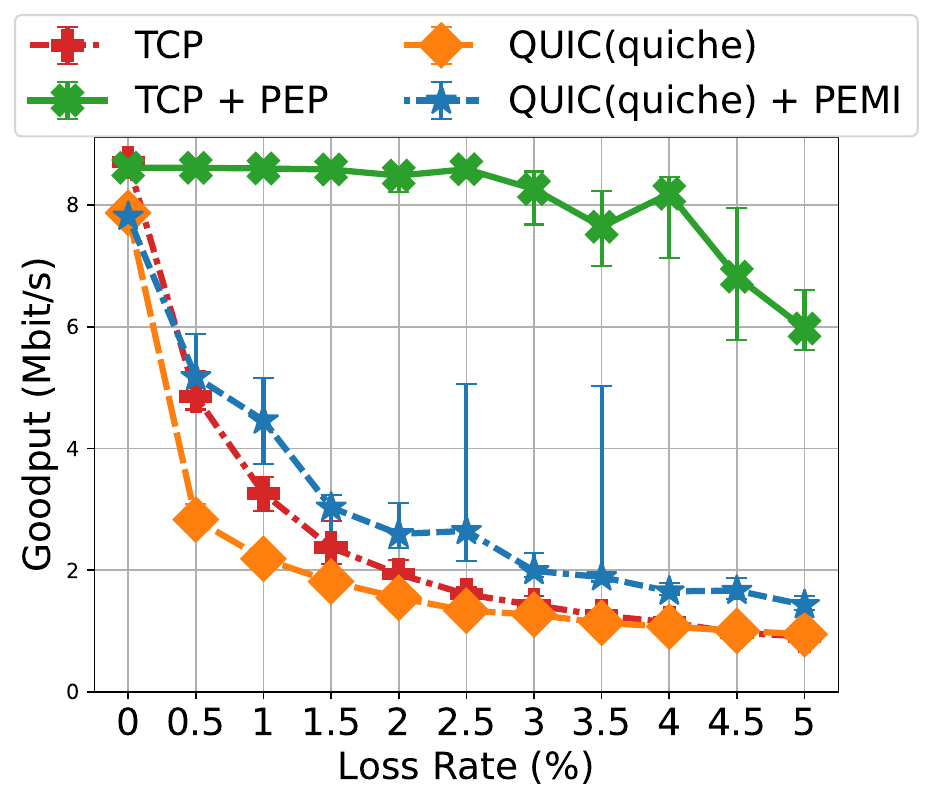}
    \caption{Goodput under different loss rates.}
    \Description{xxx.}
    \label{fig:fairness}
\end{minipage}
\end{figure}

We evaluate the performance of \name in reducing the delays of real-time communication (RTC) frames. 
The server and client are the dummy applications for RTC frames transmission~(\cref{sec:app_rtc}).
We run each time of experiment for 20 seconds, and repeat 10 times.
As illustrated in \cref{fig:rtc_delay}, \name significantly reduces both frame-level jitter and frame delay across all three QUIC stacks.
(1) Frame-level jitter~(\cref{fig:rtc_delay_jitter}). 
\name substantially reduces jitter at both the median and tail.
Across the three stacks, the median jitter is reduced by 45\%-70\%, and the tail jitter (90th-percentile) is reduced by 20\%--75\%;
(2) Frame delay~(\cref{fig:rtc_delay_frame}). 
\name also substantially reduces the frame delay for the three stacks, with the median frame delay reduced by 1\%-94\% and the tail (90th-percentile) frame delay reduced by about 59\%-87\%.

\subsection{Goodput under Different Packet Loss Rates}
\label{sec:eval_fairness}
To evaluate the aggressiveness of different schemes, we test their performance under various packet loss rates. The results are shown in \cref{fig:fairness}. 
At zero packet loss rate, all schemes performed well. 
As the packet loss rate increased, both QUIC and TCP experienced obvious performance degradation.
After being enhanced by PEP, the performance of TCP remains almost consistent under moderate packet loss. This is because PEP divides an end-to-end connection into two segments. The new connection on Link1 recovers bandwidth quickly from packet loss due to its low latency, resulting in minimal perception of packet loss by the actual sender.
Conversely, the \name-enhanced QUIC, although consistently achieves roughly twice the goodput of QUIC across all loss, shows significantly lower goodput than that of PEP-enhanced TCP. 
\name does not completely shield end-to-end connections from packet loss awareness.
\name cannot detect and retransmit all lost packets as effectively as PEP. Even for retransmitted packets, endpoints may still perceive packet loss due to out-of-order replies~\cite{rfc9002}.
Consequently, QUIC with \name exhibits less aggressive behavior than PEP-enhanced TCP.
This indicates that \name does not make QUIC non-TCP-friendly; instead, the moderate optimization from \name can mitigate the disadvantage faced by QUIC when compared to enhanced TCP, promoting fairness in network traffic.

\subsection{Performance under trace-driving dynamic networks using CellReplay}
\label{sec:eval:mahimahi}
\begin{figure}[hbpt]
    \subfloat[Frame-level jitter.]{\includegraphics[width=0.35\linewidth]{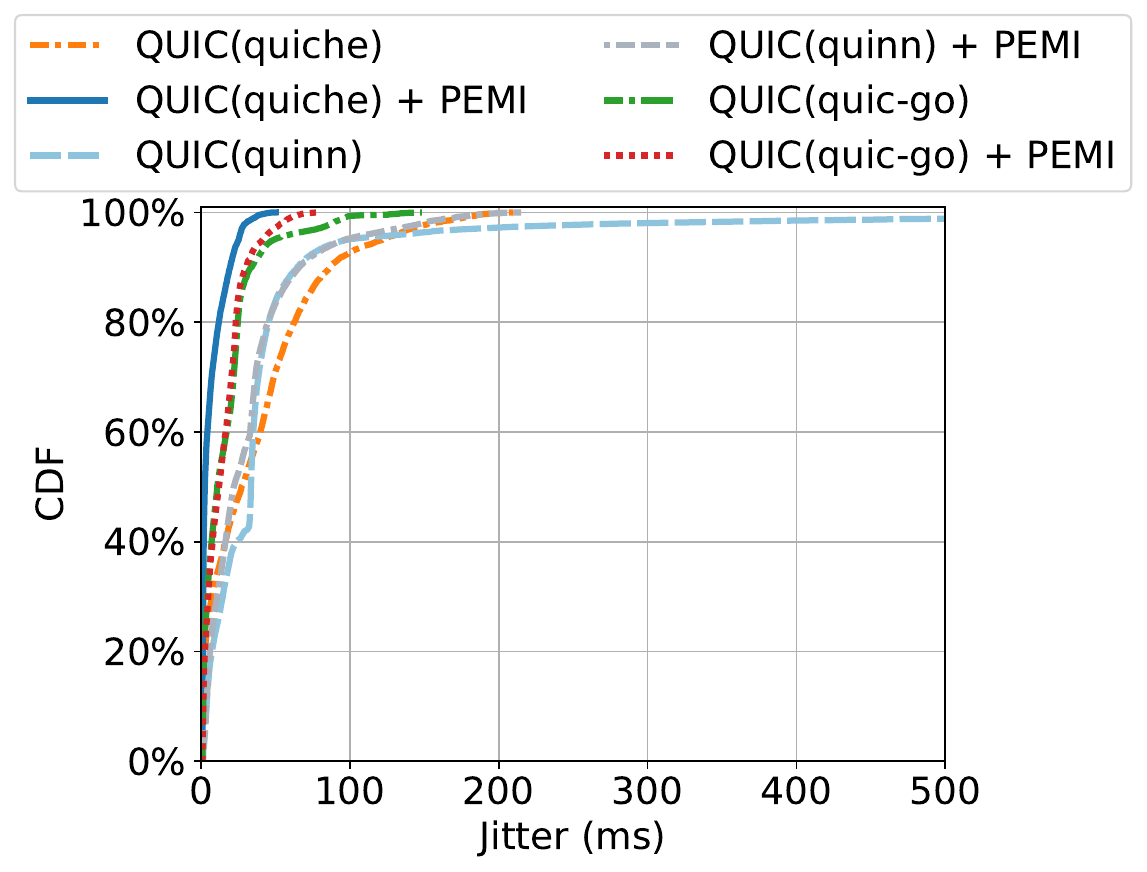}\label{fig:cell_jitter}}
    \subfloat[Frame delay.]{\includegraphics[width=0.35\linewidth]{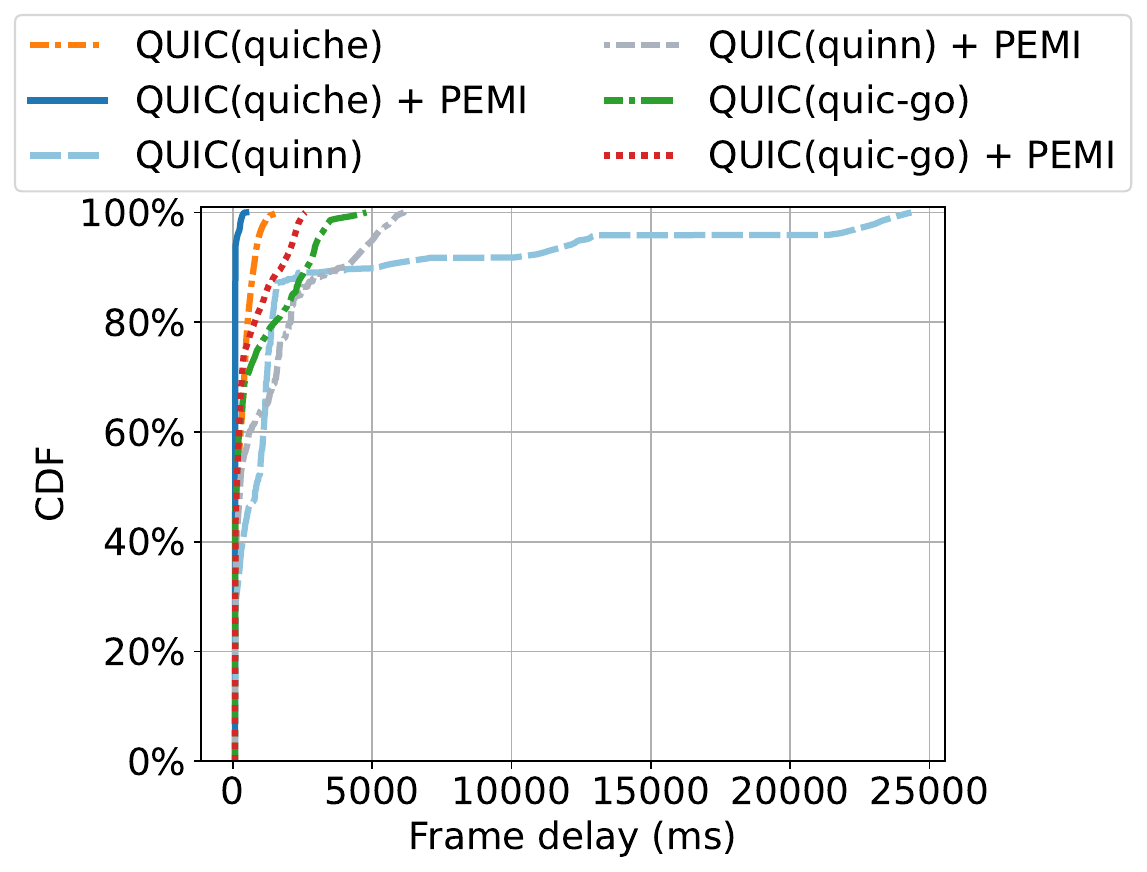}\label{fig:cell_frame_delay}}
    \caption{Performance of RTC frames transmission under CellReplay-based dynamic network.}
    \Description{xxx.}
    \label{fig:mahimahi_cell}
\end{figure}

We run the RTC frames transmission experiments under CellReplay-based environment for 30 seconds, repeating 5 times.
The results are shown in \cref{fig:mahimahi_cell}.
With \name enabled, QUIC is more resilient to burst losses in dynamic networks. 
(1) Frame-level jitter.
Enabling \name substantially reduces frame-level jitter for quiche and quinn, with median jitter reduced by 35\%--90\% and tail jitter (99th percentile) reduced by approximately 70\%--80\%.
For quic-go, while the median jitter slightly increases, the tail jitter is still significantly reduced (about 37\%);
(2) Frame delay.
\name consistently reduces frame delay for quiche and quinn, achieving 45\%--70\% reductions at the median and over 70\% reductions at the tail (99th percentile).
For quic-go, although the median frame delay slightly increases, the tail delay is substantially reduced by around 40\%.

\subsection{Computation Overhead}
\label{sec:eval_cpu}
To investigate the CPU usage of \name, we statistic CPU cycles used for I/O processing and other computation tasks, while transferring varying sizes of flows.

\begin{wrapfigure}{r}{0.44\textwidth}
        \vspace{-0.5cm}
        \centering
        \includegraphics[width=0.9\linewidth]{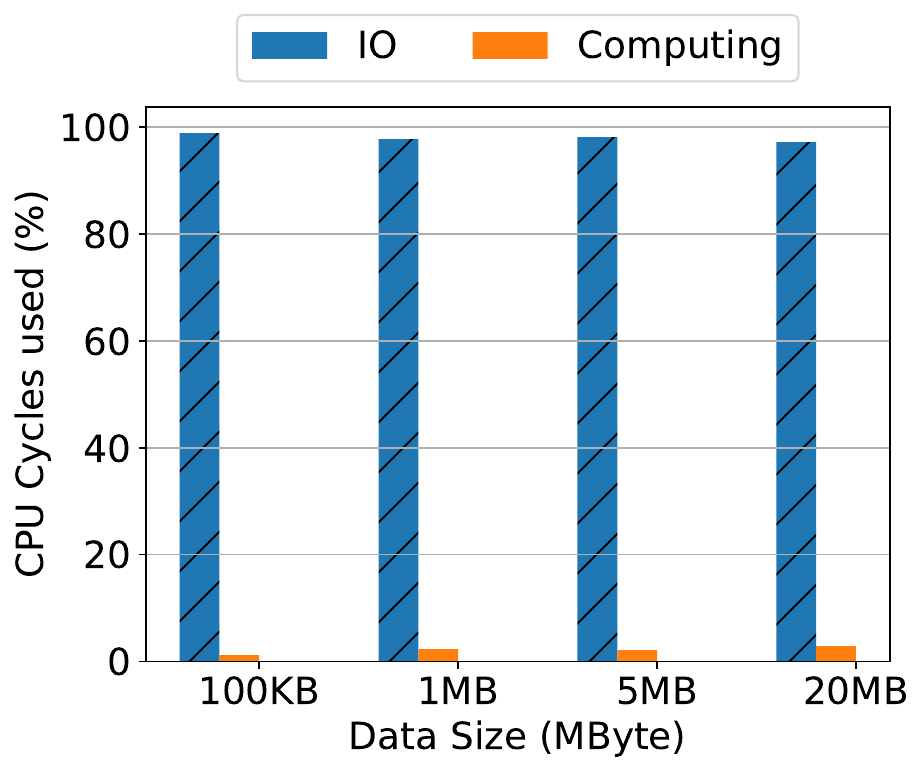}
        \caption{CPU cycles usage of \name.}
        \Description{xxx.}
        \label{fig:cpu}
\end{wrapfigure}

The results are shown in \cref{fig:cpu}. It illustrates that most of the CPU cycles are consumed by I/O processing, consisting of packet receiving and sending, which is also necessary for traditional PEP to enhance TCP traffic.
The computation introduced by the \name design, including handshake tracking, flowlet recording, loss detection, RTT measurement, and congestion control, consistently consumes less than 3\% of the CPU cycles.

\section{Limitations}
\label{sec:limitations}

Compared to PEP, \name exhibits several limitations that we briefly summarize here.

\noindent\textbf{Actively generating ACKs for congestion control.}
PEP splits an end-to-end connection into multiple connections, enabling each segment to receive ACK feedback more quickly. 
This may accelerate congestion control reactions, such as CWND growth.
In contrast, \name preserves end-to-end connections and does not actively generate ACKs, and therefore cannot directly provide such acceleration.

\noindent\textbf{Bidirectional transfer scenarios.}
When both endpoints of a QUIC connection concurrently transmit substantial amounts of data, ACK frames are frequently piggybacked on data packets. This results in an ACK pattern that is fundamentally different from that of unidirectional traffic. Under such conditions, the current mechanisms in \name may become ineffective, which is precisely why PEMI needs to detect the dominant traffic direction.

\noindent\textbf{More Complex ACK Behaviors.}
\name relies on relatively regular ACK behaviors to infer packet loss. Its current detection logic assumes that the receiver promptly returns one ACK packet for every one or two received data packets. 
However, the ACKs may be irregular or sparse.
For example, a receiver could intentionally delay ACK generation for an arbitrary period---as long as the delay does not exceed the advertised \texttt{max\_ack\_delay}~\cite{rfc9000}---and such behavior is permitted by the QUIC standard.
Further, some research efforts aim to reduce the number of ACKs~\cite{tack2020sigcomm,custura2023quicackreducing}.
\name inherently relies on the regularity of ACKs to detect packet loss. Therefore, if some connections send ACKs in a particularly low or irregular frequency, \name is unable to provide optimization for them.

\noindent\textbf{Changes in packet intervals.}
Micro-scale network jitter can cause variation in the \name observed reply timestamps of packets within a flowlet, which degrades \name's heuristic for locating partial losses that assumes per-packet delays are approximately uniform in a flowlet. For example, some queue-management schemes may cause replies for a single flowlet to arrive in rapid succession, making it difficult to associate each reply with its corresponding sent packet.

\section{Discussion}

\noindent\textbf{Middleboxes operations and ossification.}
Middleboxes are ubiquitously deployed in networks and provides crucial functions, such as network accessing, routing, firewalls, and load balancing.
They indeed occupy advantage points in addressing some performance issues like loss and congestion~\cite{sidekick2024nsdi,tecc2024nsdi}.
Therefore, research and practice in identifying and processing connections on middleboxes will continue to exist~\cite{rfc9312}.
\name adheres to two principles crucial to avoid or minimize protocol ossification:
(1) Minimize recognition of the protocol format. 
(2) Non-destructive. The middlebox solutions should avoid destructive behavior, such as modifying packets or blocking unrecognized connections.

These principles are also recommended by the IETF community~\cite{rfc9312}.
\name utilizes only the minimized fields of non-encrypted information; it does not modify any packets; and it forwards unrecognized traffic normally.
To provide a concrete example for ossification avoidance, \name is the first middlebox enhancement for QUIC that seamlessly supports MPQUIC.
TECC~\cite{tecc2024nsdi} edited QUIC frames format and parses frames at the tunnel server, thus requires modifications to support MPQUIC; Sidekick~\cite{sidekick2024nsdi} requires at least modifications to support quACK over different paths.

\noindent\textbf{Security and privacy.}
\name does not increase the data visibility of the middlebox, thus not introducing additional privacy concerns. However, a potential concern is, since \name performs retransmissions based on inference, there could be a risk of amplification attacks.
Therefore, \name retransmits only once and for no more than 10\%~(\cref{sec:best_effort}). This limits the potential amplification factor to be less than 1.1X. For comparison, QUIC restricts the amount of data the server can send to not exceed 3X before verifying the client.

\noindent\textbf{Support of other transport protocols.}
\name can be extended to other secure transport protocols if the middleboxes are able to: 
(1) measure initial RTT measurements, such as partial visibility of the header during the handshake phase; 
(2) distinguish packets belonging to a specific connection. 
Overall, the requirements of \name for protocols are minimal, requiring no visibility of ACK information, the ability to interrupt or proxy connections, and modifications to the endpoints.

\section{Conclusion}

In this paper, by analyzing the QUIC standard and stacks, and drawing inspiration from the locality of application traffic, we identify opportunities for middleboxes to transparently measure RTT and infer packet loss for QUIC connections.
We then present \name as a concrete design that demonstrates the feasibility of this
approach.
To the best of our knowledge, \name represents the first attempt to achieve this "Impossible Mission".
Without viability on plaintext ACK information, \name achieves best-effort performance enhancement through best-effort protocols (e.g., UDP) and best-effort information (e.g., spin bit).
\name leverages the locality of traffic to avoid error propagation and continuously measuring RTT and infers packet loss.
We implement a prototype of \name and extensive experimental results show that \name can significantly optimize QUIC in the presence of packet loss. 
We hope this first step could inspire more researchers to join this field, realizing transparent enhancements for QUIC / HTTP/3 and benefiting applications.

\bibliographystyle{ACM-Reference-Format}
\bibliography{paper}

\appendix

\section{Implementation}
\label{sec:impl}

We implement \name prototype in Rust with about 3k LOC. 
The primary mechanisms of \name are detailed in the \cref{sec:design}.
In this section, we introduce some other aspects related to the \name prototype.

\noindent\textbf{Packets forwarding.}
\name relies on \texttt{TPROXY} of Linux to intercept the UDP packets. \name receives the UDP packets from the \texttt{TPROXY}, and then finds the connection it belongs to by the four-tuple (IP and port at both ends).
For a UDP packet that does not belong to recognized connections, \name tries to parse it as the QUIC handshake packet to find new connections. If it is the Initial packet of a new connection, \name creates the connection and forwards the packets. For the rest UDP packets that cannot be parsed as QUIC packets, \name simply forwards them.

\begin{figure}[htpb]
    \centering
    \includegraphics[width=0.8\linewidth]{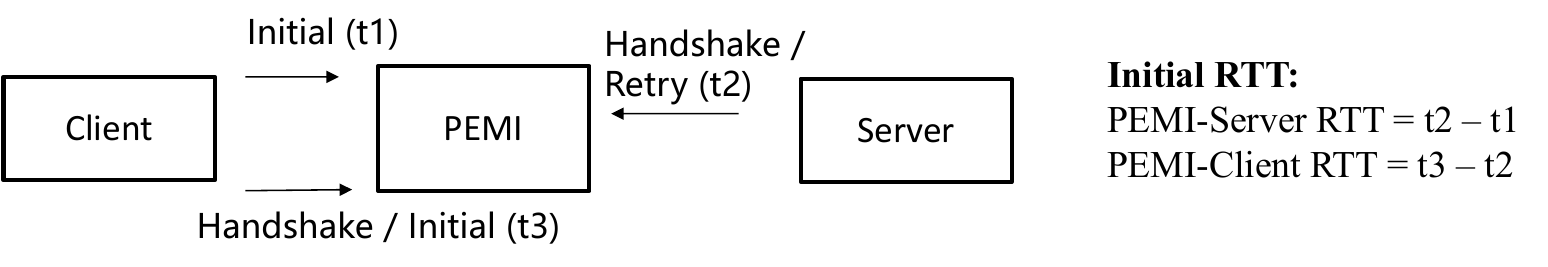}
    \caption{QUIC Initial RTT measurement.}
    \label{fig:impl_init_rtt}
\end{figure}

\noindent\textbf{QUIC connection tracking and initial RTT measurement.}
QUIC clients and servers use long header packets for the handshake, which can be used to identify new connections~\cite{rfc9312}.
It also enables \name to measure the initial RTT.
As shown in \cref{fig:impl_init_rtt}, \name parses the long header packets to find new client request (1st Initial Packet), server reply (1st Handshake Packet or a Retry Packet), and client acknowledgment (1st Handshake Packet from client or Initial Packet after a server Retry Packet). After the handshake process, \name uses the four-tuple as the connection identifier. We set an idle timeout for every connection and remove the connection if it is idle. The idle timeout is set to 120 seconds as required by~\cite{rfc9312}.
Using four-tuples as identifiers cannot support connection migration.
We have not used QUIC connection IDs because they cannot guarantee uniqueness in the Internet, so there may be more than one connection with the same ID (although the probability is low), necessitating conflict handling. We leave using connection IDs for future implementation exploration. 

\noindent\textbf{Measuring the Eliciting Threshold.}
\name continuously estimates the \texttt{eliciting\_threshold} based on the ratio between sent packets and reply packets. If more than 60\% of sent packets receive replies, the threshold is inferred to be~1; otherwise, it is inferred to be~2. The 60\% cutoff is chosen based on empirical observations. As shown in \cref{sec:quicackstudy}, mainstream QUIC implementations exhibit clearly separated reply-to-sent ratios, making this heuristic insensitive in practice.
In our evaluation, the inferred threshold converges quickly—typically within tens to hundreds of packets—and remains stable throughout the lifetime of a connection.

\noindent\textbf{RTT Calibration.}
Because the QUIC stacks(quiche, quic-go, quinn) used in our evaluation (\cref{sec:eval}) do not support the spin bit, the current implementation of \name performs RTT calibration using ICMP probes. 
We utilize one ICMP-based \name-Receiver RTT sample every end-to-end RTT, which provides the same sampling frequency as the spin bit.
Therefore, the results reported in this paper are expected to be similar to those obtained when spin-bit support is available.
When the deviation of the RTT exceeds the threshold $\Delta_{\text{margin}}$ (defined in \cref{eq:flowlet_reply_find}), the RTT samples inferred from flowlets are treated as unreliable by \name.
In such cases, we reset the measured RTT using the ICMP-based RTT sample and clear all flowlets that have already been matched with replies, in order to prevent incorrect RTT samples from contaminating subsequent measurements.

\section{Related Work}

\smalltitle{Middleboxes helped congestion control.}
In addition to PEP, there is another kind of in-network assistance that utilizes information from middleboxes to optimize end-to-end congestion control~\cite{abc2020nsdi,zhuge2022sigcomm}. ABC~\cite{abc2020nsdi} detects congestion conditions on routers and feeds this information back to the endpoints to adjust their transmission strategies. Zhuge~\cite{zhuge2022sigcomm} predicts delays through queuing information on Wi-Fi access points, and feedbacks to sender-side. Compared to PEP/\name, neither of these solutions can assist loss recovery by rapid retransmissions.
Furthermore, ABC requires concurrent modifications to both the end host and the middlebox; Zhuge is only deployable at exact congestion points with big queues, since it requires access to the queue to predict the delay.

\smalltitle{Performance enhancement for QUIC.}
A few recent works attempt to optimize the transmission performance of QUIC with the help of middleboxes~\cite{tecc2024nsdi,sidekick2024nsdi}. TECC~\cite{tecc2024nsdi} establishes a tunnel between end-to-end QUIC connections and sets the target speed of the endpoint server based on the available bandwidth detected by the tunnel server. The tunnel server also provides retransmission information to avoid duplicate retransmissions from the end. 
Sidekick~\cite{sidekick2024nsdi} transmits quACKs from middleboxes through a dedicated protocol.
QuACKs provide packet reception information visibility between endpoints and middleboxes.
It modifies the QUIC end to utilize quACKs and enables performance enhancements like fast retransmissions. 
Sidekick is able to achieve optimization effects comparable to PEPs~\cite{sidekick2024nsdi}.
These works are fundamentally different from TCP-era PEPs in terms of transparency: both TECC and Sidekick require simultaneous modification of the end hosts and middleboxes, and newly defined interaction messages between hosts and middleboxes during runtime.
In contrast, \name requires deployment only on middleboxes, thereby maintaining transparency with respect to the endpoints.

\section{Artefact Availability}
\label{sec:appendix:opensource}

We open-source the \name prototype implementation and related research tools on GitHub at \url{https://github.com/zhjie233/pemi}.
The tools enable reproducible studies of performance enhancement mechanisms' effectiveness and performance.
We hope this could facilitate the future researches in this area.
Specifically, the tools integrated with the \name prototype provide the following capabilities:
\begin{enumerate}
    \item Collecting traces from the perspective of middleboxes. Using the \name prototype, we record the arrival time, direction, and size of each packet observed by the middlebox.
    \item Obtaining ground truth of packet loss and latency for mechanisms evaluation. Due to QUIC's encrypted ACKs, this requires combining information from both the endpoints and middleboxes. Specifically, the tools analyze (1) packet capture and (2) log analysis from both the endpoints and \name.
\end{enumerate}

For ground truth acquisition, we recommend using logs as the information source, as we found that packet capture may miss packets, leading to misjudgment of packet loss.
To track packet lifetime by logs, we use the same function (first 8 bytes + last 8 bytes of UDP payload) at both endpoints and \name to assign an ID to each packet, and then analyze which packets fail to reach the next hop.
We treat the first packet sent by the receiver after receiving a data packet as the reply to that data packet. By correlating their timestamps observed at the \name, we can calculate the RTT ground truth between the middleboxes and the endpoints.

\end{document}